\documentstyle[12pt]{article}
\newcommand{\too}{\longrightarrow}

\newcommand{\T}{{\cal T}}

\newcommand{\s}{{\cal S}}

\newcommand{\D}{{\cal D}}

\newcommand{\p}{{\cal P}}

\newcommand{\di}{\displaystyle}
\newcommand{\Om}{\Omega}
\newcommand{\na}{\nabla}

\newcommand{\al}{\alpha}

\newcommand{\Ga}{\Gamma}

\newcommand{\la}{\lambda}
\newcommand{\De}{\Delta}
\newcommand{\de}{\delta}

\def \reel{ {\rm I}\!{\rm R} }

 \def \comp{ \;{}^{ {}_\vert }\!\!\!{\rm C} }

 \def \nat{ { {\rm I}\!{\rm N}} }

 \def \rat{ {\rm Q}\kern-.65em {}^{{}_/ }}

\newtheorem{th}{Theorem}[section]

\newtheorem{Le}{Lemma}[section]

\newtheorem{rem}{Remark}[section]
\title{ Spectra and symmetric eigentensors of the Lichnerowicz Laplacian
 on $P^n(\comp)$ }
\author{Mohamed Boucetta \footnote{Recherche
men\'ee dans le cadre du Programme
 Th\'ematique d'Appui \`a la Recherche Scientifique PROTARS III.}} \date{ }\parindent=0cm
\begin{document}
\maketitle

{\bf Abstract.} We compute the eigenvalues with multiplicities  of
the Lichnerowicz Laplacian acting on
 the space of complex symmetric covariant
tensor fields on the complex projective space $P^n(\comp)$. The
spaces  of symmetric eigentensors are explicitly given.

\bigskip



\section{Introduction}

 Let $(M,g)$ be a Riemannian $n$-manifold and $D$  its Levi-Civita connection. For
any $p\in\nat$, we shall denote by $\Ga(\otimes^pT^*M,\comp)$,
$\Om^p(M,\comp)$ and $\s^p(M,\comp)$ the space of complex
covariant $p$-tensor fields on $M$, the space of complex
differential $p$-forms on $M$ and the space of complex symmetric
covariant $p$-tensor fields on $M$, respectively. Note that
$\Ga(\otimes^0T^*M,\comp)=\Om^0(M,\comp)=\s^0(M,\comp)=C^\infty(M,\comp)$.
We put
$$\di\Om(M,\comp)=\bigoplus_{p=0}^n\Om^p(M,\comp)\quad\mbox{and}\quad
\s(M,\comp)=\bigoplus_{p\geq0}\s^p(M,\comp).$$  If $(M,g)$ is
K\"ahlerian, i.e., there exists a complex structure $J$ on $M$
such that $DJ=0$ and $g$ is Hermitian with respect to $J$. The
complex structure $J$ defines a bigraduation
$$\Om^p(M,\comp)=\bigoplus_{r+q=p}\Om^{r,q}(M)\quad\mbox{and}\quad
\s^p(M,\comp)=\bigoplus_{r+q=p}\s^{r,q}(M).$$

We consider the Lichnerowicz Laplacian
$\De_M:\Ga(\otimes^{*}T^*M,\comp)\too\Ga(\otimes^{*}T^*M,\comp)$
 introduced by Lichnerowicz in [16] pp. 26. It  is a second order differential operator,
 self-adjoint,
elliptic and respects the symmetries of tensor fields. In
particular, $\De_M$ leaves invariant $\s(M,\comp)$
 and the restriction of
$\De_M$ to  $\Om(M,\comp)$ coincides with the Hodge-de Rham
Laplacian. Moreover, the Lichnerowicz Laplacian respects the
bigraduation induced by   K\"ahlerian structures.

The Lichnerowicz Laplacian  acting on symmetric covariant tensor
fields is of fundamental importance in mathematical physics  (see
for instance  [10], [21] and [23]). Note also that the
Lichnerowicz Laplacian acting on symmetric covariant 2-tensor
fields appears in many problems in Riemannian geometry (see [3],
[5], [20]...).

On a compact Riemannian  manifold, the Lichnerowicz Laplacian
$\De_M$ has discrete eigenvalues with finite
 multiplicities.
 For a given compact Riemannian  manifold, it may be an interesting problem
 to determine explicitly
the eigenvalues and the eigentensors of $\De_M$ on $M$.

Let us enumerate the cases where the spectra of $\De_M$ was
computed:
\begin{enumerate}
\item $\De_M$ acting on $C^\infty(M,\comp)$:  $M$ is either flat
torus or
 Klein bottles [4], $M$ is a Hopf manifold
 [1];
\item $\De_M$ acting on  $\Om(M,\comp)$: $M=S^n$ or $P^n(\comp)$
[11] and [12], $M=\comp aP^2$ or $G_2/SO(4)$ [17]-[19],
$M=SO(n+1)/SO(2)\times SO(n)$ or $M=Sp(n+1)/Sp(1)\times Sp(n)$
[22]; \item $\De_M$ acting on $\s^2(M,\comp)$ and $M$ is the
complex projective space $P^2(\comp)$ [23]; \item $\De_M$ acting
on $\s^2(M,\comp)$ and $M$ is either $S^n$ or
 $P^n(\comp)$ [6] and [7];
 \item  Brian and Richard Millman gave in [2] a theoretical
 method for
computing the spectra of Lichnerowicz Laplacian acting on $\Om(G)$
where $G$ is a compact semisimple Lie group endowed with the
biinvariant metric induced from the negative of the Killing form;
\item Some partial results where given in [13]-[15]; \item $\De_M$
acting on $\s(M,\comp)$ and $M$ is $S^n$ $[8]$.\end{enumerate}

In this paper, we compute the eigenvalues and we determine the
spaces of eigentensors with its multiplicities of $\De_M$ acting
on $\s(M,\comp)$ in the case where $M$ is the complex projective
space $P^n(\comp)$ endowed with the K\"ahlerian structure quotient
of the canonical K\"ahlerian structure of $\comp^{n+1}$. We use a
fairly simple method which requires, in places, massive
computations. Let us describe this method briefly.

First, since there is a natural map
$$\phi:\Ga_Z(\otimes^*T^*\comp^{n+1},\comp)\too\Ga(\otimes^*T^*P^n(\comp),\comp),$$where
$\Ga_Z(\otimes^*T^*\comp^{n+1},\comp)$ is the space of complex
covariant tensor fields invariant by the natural action of the
circle on $\comp^{n+1}$, we compute
$$\phi\circ\De_{\comp^{n+1}}-\De_{P^n(\comp)}\circ\phi.$$ The
formula obtained (cf. Theorem 2.1) involves natural operators on
$\comp^{n+1}$ and constitutes the principal  tool of this paper.
Hereafter, in Section 3, we adapt  to our situation the methods
used in [11] in the context of differential forms. Indeed, we
consider, for any $p,q,k,l\in \nat$, the space $T_{k,l}^{p,q}$ of
traceless symmetric tensor field $T$ on $\comp^{n+1}$ of the form
$$T=\sum_{\stackrel{0\leq i_1<\ldots,<i_p\leq n}{0\leq j_1<\ldots,<j_q\leq
n}}T_{i_1,\ldots,i_p,j_1,\ldots,j_q}dz_{i_1}\odot\ldots\odot
dz_{i_p}\odot d\bar z_{j_1}\odot\ldots\odot d\bar z_{j_q},$$where
$T_{i_1,\ldots,i_p,j_1,\ldots,j_q}$ are harmonic polynomials of
degree $k$ with respect $z_0,\ldots,z_n$ and of degree $l$ with
respect $\bar z_0,\ldots,\bar z_n$ and such that the divergence of
$T$ vanishes. We have, if $<\;,\;>$ denotes the Euclidian metric
on $\comp^{n+1}$,
$$\di\phi:\bigoplus_{\stackrel{0\leq
m\leq\min(p,q)}{k+p=l+q}}<\;,\;>^m\odot
T_{k,l}^{p-m,q-m}\too\s^{p,q}(P^n(\comp),\comp)$$ is injective and
its image is dense. By introducing  an algebraic lemma (cf. Lemma
3.3) we get a direct sum decomposition of any $T_{k,l}^{p,q}$, and
we use the formula obtained in Theorem 2.1 to show that the image
by $\phi$ of the spaces composing  this direct sum are, actually,
eigenspaces of $\De_{P^n(\comp)}$.  We compute the multiplicities
of these eigenspaces and we get the result desired (see Theorems
3.2 and 3.3).  Finally,, we tabulate the results for the low
values of $p$ and $q$ (Tables I-VIII) and, in particular, we
recover the results obtained in [23] (Tables VI-VIII).

\section{A relation between $\De_{\comp^{n+1}}$ and $\De_{P^n(\comp)}$}

The main result of this section is Theorem 2.1 which establishes a
formula  relating the Lichnerowicz Laplacian on $\comp^{n+1}$ and
the Lichnerowicz Laplacian on $P^n(\comp)$. This formula is the
principal tool of this paper and its statement requires the
introduction of some definitions and notations. Also we need to
recall some basic properties of the Lichnerowicz Laplacian and to
collect the basic material which will be used throughout the
paper.

\bigskip

Let $(M,g)$ be a Riemannian $n$-manifold. The curvature tensor
field $R$ of  the Levi-Civita connection $D$ associated to $g$
 is given by
\[R(X,Y)Z=D_{[X,Y]}Z-\left(D_XD_YZ-D_YD_XZ\right),\]and
its Ricci endomorphism field $r:TM\too TM$
  is given by
\[g(r(X),Y)=\sum_{i=1}^ng(R(X,E_i)Y,E_i),\]where
$(E_1,\ldots,E_n)$ is any local orthonormal frame.

 For any $p\in\nat$, the connection $D$ induces a
differential operator
$$D:\Ga(\otimes^pT^*M,\comp)\too\Ga(\otimes^{p+1}T^*M,\comp)$$
given by
\[DT(X,Y_1,\ldots,Y_p):=D_XT(Y_1,\ldots,Y_p)=X.T(Y_1,\ldots,Y_p)-
\sum_{j=1}^pT(Y_1,\ldots,D_XY_j,\ldots,Y_p).\]  Its formal adjoint
$D^*:\Ga(\otimes^{p+1}T^*M,\comp)\too\Ga(\otimes^{p}T^*M,\comp)$
is given by
\[D^*T(Y_1,\ldots,Y_p)=-\sum_{j=1}^nD_{E_i}T(E_i,Y_1,\ldots,Y_p),\]
where $(E_1,\ldots,E_n)$ is any local orthonormal frame.

We  denote by $\de$  the restriction of $D^*$ to $\s(M,\comp)$ and
we define $\de^*:\s^p(M,\comp)\too\s^{p+1}(M,\comp)$ by
$$\de^*T(X_1,\ldots,X_{p+1})=\sum_{j=1}^{p+1}D_{X_j}T(X_1,\ldots,
\hat{X}_j,\ldots,X_{p+1}),$$ where the symbol $\;\hat{}$ means
that the term is omitted.

 Recall that the operator trace
${\mathrm{Tr}}:\s^p(M,\comp)\too\s^{p-2}(M,\comp)$ is given by
$${\mathrm{Tr}}T(X_1,\ldots,X_{p-2})=\sum_{j=1}^nT(E_j,E_j,X_1,\ldots,X_{p-2}),$$
where $(E_1,\ldots,E_n)$ is any local orthonormal frame.

The Lichnerowicz Laplacian is the second order differential
operator
$$\De_M:\Ga(\otimes^{p}T^*M,\comp)\too\Ga(\otimes^{p}T^*M,\comp)$$ given
by
\[\De_M(T)=D^*D(T)+R(T),\] where $R(T)$ is the curvature
operator given by
\begin{eqnarray*}
&&R(T)(X_1,\ldots,X_p)=\sum_{j=1}^pT
(X_1,\ldots,r(X_j),\ldots,X_p)\\
&-&\sum_{i<j}\sum_{l=1}^n\left\{T(X_1,\ldots,E_l,\ldots,R(X_i,E_l)X_j,\ldots,X_p)
+
T(X_1,\ldots,R(X_j,E_l)X_i,\ldots,E_l,\ldots,X_p)\right\},\nonumber\end{eqnarray*}
where $(E_1,\ldots,E_n)$ is any local orthonormal frame and, in
$$T(X_1,\ldots,E_l,\ldots,R(X_i,E_l)X_j,\ldots,X_p),$$ $E_l$ takes
the place of $X_i$ and $R(X_i,E_l)X_j$ takes the place of $X_j$.

This differential operator, introduced by Lichnerowicz in [16] pp.
26, is self-adjoint, elliptic and respects the symmetries of
tensor fields. In particular, $\De_M$ leaves invariant
$\s(M,\comp)$
 and the restriction of
$\De_M$ to  $\Om(M,\comp)$ coincides with the Hodge-de Rham
Laplacian.

Note that if $T\in\s(M)$ then
\begin{eqnarray}{\mathrm{Tr}}(\De_M T)&=&\De_M( {\mathrm{Tr}} T),\\
\De_{M}(T\odot g)&=&(\De_MT)\odot g,\end{eqnarray}where $\odot$ is
the symmetric product.

The Lichnerowicz Laplacian is compatible with  K\"ahlerian
structures. Indeed, suppose that $(M,g)$ is K\"ahlerian, i.e.,
there exists a complex structure $J$ on $M$ such that $DJ=0$ and
$g$ is Hermitian with respect to $J$. The complex structure $J$
defines a bigraduation
$$
\s^p(M,\comp)=\bigoplus_{r+q=p}\s^{r,q}(M),$$and $\De_M$ respects
this bigraduation.

 For any
$T\in\Ga(\otimes^pT^*M,\comp)$, for any  vector field $Y$ and for
any $1\leq
 i<j\leq p$, we denote by $i_{Y,j}T$ the
$(p-1)$-tensor field given by
\[
i_{Y,j}T(X_1,\ldots,X_{p-1})=T(X_1,\ldots,X_{j-1}, Y,X_j,\ldots,
X_{p-1}),
\]
 by ${\mathrm{Tr}}_{i,j}T$ the $(p-2)$-tensor field given by
\[ {\mathrm{Tr}}_{i,j}T(X_1,\ldots,X_{p-2})=\sum_{l=1}^{n}T(X_1,\ldots,X_{i-1},
E_l,X_i,\ldots,X_{j-2},E_l,X_{j-1},\ldots, X_{p-2}),\]and by
${\mathrm{Tr}}_{i,j,J}T$ the $(p-2)$-tensor field given by
\[ {\mathrm{Tr}}_{i,j,J}T(X_1,\ldots,X_{p-2})=\sum_{l=1}^{n}T(X_1,\ldots,X_{i-1},
E_l,X_i,\ldots,X_{j-2},JE_l,X_{j-1},\ldots, X_{p-2}),\] where
$(E_1,\ldots,E_n)$ is any local orthonormal frame of $M$.

\begin{rem} If $T$ is a complex
symmetric covariant tensor field then
$${\mathrm{Tr}}_{i,j}T={\mathrm{Tr}}T\quad\mbox{ and }\quad {\mathrm{Tr}}_{i,j,J}T=0.$$
\end{rem}

For any permutation $\sigma$ of $\{1,\ldots,p\}$, we denote by
$T^\sigma$ the $p$-tensor field
\[
T^\sigma(X_1,\ldots,X_{p})=T(X_{\sigma(1)},\ldots,X_{\sigma(p)}).\]
For $1\leq i<j\leq p$, the transposition of $(i,j)$ is the
permutation $\sigma_{i,j}$ of $\{1,\ldots,p\}$ such that
$\sigma_{i,j}(i)=j$, $\sigma_{i,j}(j)=i$ and $\sigma_{i,j}(k)=k$
for $k\not=i,j$. We shall denote by $\T$  the set of the
transpositions of $\{1,\ldots,p\}$.\bigskip

On other hand, for $p\geq2$, we denote by $T^J$ and $\overline{
T}^{J}$ the $p$-tensors fields given by
\begin{eqnarray*}
T^J(X_1,\ldots,X_p)=\sum_{i<j}T(X_1,\ldots,JX_i,\ldots,JX_j,\ldots,X_p),\\
\overline
{T}^J(X_1,\ldots,X_p)=\sum_{i<j}T(X_1,\ldots,JX_j,\ldots,JX_i,\ldots,X_p).\end{eqnarray*}

Finally,  we  define $\de^{*c}:\s^p(M,\comp)\too\s^{p+1}(M,\comp)$
by
$$\de^{*c}T(X_1,\ldots,X_{p+1})=\sum_{j=1}^{p+1}D_{JX_j}T(X_1,\ldots,
\hat{X}_j,\ldots,X_{p+1}),$$ and we put
$$\de^*_h=\frac12(\de^*-i\de^{*c})
\quad\mbox{and}\quad
\overline{\de^*_h}=\frac12(\de^*+i\de^{*c}).$$ Note  that
\begin{equation}
\de_h^*\circ\overline{\de^*_h}=\overline{\de^*_h}\circ\de_h^*.\end{equation}
\bigskip

The complex projective space $P^n(\comp)$ inherits a natural
K\"ahlerian structure from $\comp^{n+1}$, let us describe this
structure and introduce some notations.

Let $(z_0,\ldots,z_n)$ be the standard holomorphic coordinates on
$\comp^{n+1}$. Put $z_i=x_i+\sqrt{-1}y_i$,
$$\frac{\partial}{\partial z_i}=\frac12\left(\frac{\partial}{\partial
x_i}-\sqrt{-1}\frac{\partial}{\partial
y_i}\right)\quad\mbox{and}\quad\frac{\partial}{\partial \bar
z_i}=\frac12\left(\frac{\partial}{\partial
x_i}+\sqrt{-1}\frac{\partial}{\partial y_i}\right).$$ The standard
complex structure $J_0$ of $\comp^{n+1}$ is given by
$$J_0\frac{\partial}{\partial z_i}=\sqrt{-1}\frac{\partial}{\partial
z_i}\quad\mbox{and}\quad J_0\frac{\partial}{\partial \bar
z_i}=-\sqrt{-1}\frac{\partial}{\partial \bar z_i}.$$

 Let  $<\;,\;>=\sum_{i=0}^ndz_i.d\bar z_i$ be the flat
K\"ahler metric on $\comp^{n+1}$ and let
$\Om_0=-\sqrt{-1}\sum_{i=0}^ndz_i\wedge d\bar z_i$ be its K\"ahler
form.

The radial vector field
$\overrightarrow{r}=\sum_{i=0}^{n}\left(x_i\frac{\partial}{\partial
x_i}+y_i\frac{\partial}{\partial y_i}\right)$ splits to $
\overrightarrow{r}=W+\overline{W},$ where
$$W=\sum_{i=0}^nz_i\frac{\partial}{\partial z_i}\quad\mbox{and}
\quad \overline{W}=\sum_{i=0}^n\bar z_i\frac{\partial}{\partial
\bar z_i}.$$ Put $Z=J_0\overrightarrow{r}.$

The  differential of $\di r^2=\sum_{i=0}^n|z_i|^2$ splits to $
dr^2=W^*+\overline{W}^*,$ where
$$ W_0^*=\sum_{i=0}^n\bar z_idz_i,\quad\mbox{and}\quad
\overline{W}_0^*=\sum_{i=0}^nz_id\bar z_i.$$

Let $\pi:\comp^{n+1}\setminus\{0\}\too P^n(\comp)$  be the natural
projection  and $\pi_s:S^{2n+1}\too P^n(\comp)$ its restriction to
$S^{2n+1}\subset\comp^{n+1}\setminus\{0\}$. For any $m\in
S^{2n+1}$, put $F_m=\ker((\pi_s)_*)_m$ and let $F_m^\perp$ be the
orthogonal complementary subspace to $F_m$ in $T_m(S^{2n+1})$;
$$T_m(S^{2n+1})=F_m\oplus F_m^\perp.$$We introduce the Riemannian
metric $g$ on $P^n(\comp)$ so that the restriction of $(\pi_s)_*$
to $F_m^\perp$ is an isometry onto $T_{\pi(m)}(P^n(\comp))$. The
standard complex structures $J$ on  $P^n(\comp)$ is given by $$
J(\pi_s)_*v=(\pi_s)_*J_0v,\; v\in F_m^\perp. $$

For any vector field $X$ tangent to $P^n(\comp)$, there exists an
unique vector field $X^h$ tangent to $S^{2n+1}$ satisfying, for
any $m\in S^{2n+1}$,
$$X^h(m)\in F_m^\perp\quad\mbox{and}\quad (\pi_s)_*(X^h)= X.$$The
vector field  $X^h$ is the horizontal left of $X$.

For any $p,q\in\nat$, we  denote by
$\Ga_Z(\otimes^pT^*\comp^{n+1},\comp)$ and
$\s^{p,q}_Z(\comp^{n+1})$ the space of complex covariant
$p$-tensor fields on $\comp^{n+1}$ and the space of complex
symmetric covariant tensor fields of type $(p,q)$ on
$\comp^{n+1}$, respectively, which are invariant by $Z$. A tensor
field $T$ belongs to $\Ga_Z(\otimes^pT^*\comp^{n+1})$ if and only
if $L_ZT=0$.

We define a linear map
$$\phi:\Ga_Z(\otimes^pT^*\comp^{n+1},\comp)\too\Ga(\otimes^pT^*P^n(\comp),\comp)$$by
$$\phi(T)(X_1,\ldots,
X_p)(\pi_s(m))=T(X^h_1,\ldots,X^h_p)(m),\quad m\in S^{2n+1}.$$ The
map $\phi$ is well defined and
$\phi\left(\s^{p,q}_Z(\comp^{n+1})\right)\subset\s^{p,q}(P^n(\comp),\comp)$.

Note that  the K\"ahler form $\Om$ of $P^n(\comp)$ satisfies
$\Om=\phi(\Om_0)$.\bigskip

The Lichnerowicz Laplacian on $P^n(\comp)$ involves the curvature
operator and we will compute it now.
\begin{Le}The tensor curvature $R$ and the Ricci endomorphism field $r$ associated
to the Riemannian metric $g$ on $P^n(\comp)$ are given
by\begin{eqnarray*}
R(X_1,X_2)X_3&=&g(X_1,X_3)X_2-g(X_2,X_3)X_1-2g(JX_2,X_1)JX_3+g(JX_3,X_2)JX_1\\&&-g(JX_3,X_1)
JX_2\\
r(X)&=&2(n+1)X.\end{eqnarray*}\end{Le}

{\bf Proof.} These formulas can be deduced easily from the
curvature of $S^{2n+1}$ by using the Riemannian submersion
$\pi_s:S^{2n+1}\too P^n(\comp)$ and the O'Neil formulas (see for
instance [5, pp. 241]).$\Box$\bigskip

A direct computation using Lemma 2.1 and the definition of the
curvature operator gives the following lemma.

\begin{Le}For any covariant
$p$-tensor field $T$ on $P^n(\comp)$, we have $$\begin{array}{lll}
R(T)(X_1,\ldots,X_p)&=&2(n+1)pT(X_1,\ldots,X_p)-4T^J(X_1,\ldots,X_p)
\\
&&\di-2\overline{T}^J(X_1,\ldots,X_p)
+2\sum_{\sigma\in\T}T^\sigma(X_1,\ldots,X_p)\\
&&\di-2\sum_{i<j}g(X_i,X_j){\mathrm{Tr}}_{i,j}T(X_1,\ldots,\hat
X_i,\ldots,\hat X_j,\ldots,X_p)
\\
&&\di-2\sum_{i<j}\Om(X_i,X_j){\mathrm{Tr}}_{i,j,J}T(X_1,\ldots,\hat
X_i,\ldots,\hat X_j,\ldots,X_p).
\end{array}$$\end{Le}

Now we are  able to state the main result of this section.

\begin{th} Let $T\in\Ga_Z(\otimes^pT^*\comp^{n+1},\comp)$. Then
\begin{eqnarray*}
\phi(\De_{\comp^{n+1}}T)-\De_{P^n(\comp)}\phi(T)&=&\phi\left(p(1-p)T+2(p-n)
L_{\overrightarrow{r}}T-L_{\overrightarrow{r}}\circ
L_{\overrightarrow{r}}T\right.\\
&&\left.-2\sum_{\sigma\in\T}T^\sigma+2T^{J_0}+2\overline{T}^{J_0}+O(T)\right),\end{eqnarray*}where
\begin{eqnarray*}
&&O(T)(X_1,\ldots,X_p)=\\
&&+2\sum_{j=1}^p\left(D_{J_0X_j}(i_{J_0\overrightarrow{r},j}T)(X_1,\ldots,
\hat{X_j},\ldots,X_p)
-D_{X_j}(i_{\overrightarrow{r},j}T)(X_1,\ldots,
\hat{X_j},\ldots,X_p)\right)\\
&&+2\sum_{i<j}<X_i,X_j>{\mathrm{Tr}}_{i,j}T(X_1,\ldots,\hat X_i
,\ldots,\hat X_j,\ldots,X_p)\\
&&+2\sum_{i<j}\Om_0(X_i,X_j){\mathrm{Tr}}_{i,j,J_0}T(X_1,\ldots,
\hat X_i ,\ldots,\hat X_j,\ldots,X_p).\end{eqnarray*}\end{th}

{\bf Proof.} The proof is a massive computation in a local
orthonormal frame.

For any vector field $X$ tangent to $P^n(\comp)$, even if its
horizontal left $X^h$ is a vector field tangent to  $S^{2n+1}$,
sometimes we need to  extent it to a local vector field on
$\comp^{n+1}$ and we continue to note it by $X^h$.

We choose a local orthonormal frame of $\comp^{n+1}$
 of the form $(E_1^h,\ldots,E_{2n}^h,N,J_0N)$ in a neighborhood of a point
 $m\in S^{2n+1}$
  such
that $(E_1^h,\ldots,E_{2n}^h)$ is the horizontal left of a local
orthonormal frame $( E_1,\ldots, E_{2n})$ of $P^n(\comp)$ in a
neighborhood of $\pi_s(m)$ and $N=\frac1r\overrightarrow{r}$ where
$r=\sqrt{|z_0|^2+\ldots+|z_n|^2}$.

Let $D$ be the Levi-Civita connection associated to the flat
Riemannian metric on $\comp^{n+1}$. For any vector field $X$ on
$\comp^{n+1}$, we have
\begin{eqnarray} D_XN&=&\frac1r\left(X-<X,N>N\right),\\
D_NX&=&[N,X]+\frac1r(X-<X,N>N),\\
D_XJ_0N&=&\frac1r(J_0X-<X,N>J_0N),\\
D_{J_0N}X&=&[J_0N,X]+\frac1r(J_0X-<X,N>J_0N).\end{eqnarray}

 Let $\na$ be the
Levi-Civita connexion of the Riemannian metric $g$ on
$P^n(\comp)$. We have, for any vector fields $ X, Y$ tangent to
$P^n(\comp)$ and in restriction to $S^{2n+1}$,
\begin{equation} D_{X^h}Y^h=(\na_{ X}Y)^h+\frac1r<J_0Y^h,X^h>J_0N-\frac1r<X^h,Y^h>N.\end{equation}

Let $T$ be a covariant $p$-tensor field on $\comp^{n+1}$ such that
$L_ZT=0$ and $(X_1,\ldots,X_p)$ a family of vector fields on
$P^n(\comp)$ in a neighborhood of $\pi_s(m)$.

A direct calculation using the definition of the Lichnerowicz
Laplacian gives
$$
\di\De_{\comp^{n+1}}(T)(X_1^h,\ldots,X_p^h)=
D^*D(T)(X_1^h,\ldots,X_p^h)=Q_1+Q_2+Q_3+Q_4,$$where
\begin{eqnarray*}
Q_1&=&\sum_{i=1}^{2n}\left(-E_i^hE_i^h.T(X_1^h,\ldots,X_p^h)
+2\sum_{j=1}^pE_i^h.T(X_1^h,\ldots,D_{E_i^h}X_j^h,\ldots,X_p^h)\right.\\
&&\di+D_{E_i^h}E_i^h.T(X_1^h,\ldots,X_p^h)
-\sum_{j=1}^pT(X_1^h,\ldots,D_{D_{E_i^h}E_i^h}X_j^h,\ldots,X_p^h)\\
&&\left.\di-\sum_{j=1}^pT(X_1^h,\ldots,D_{E_i^h}D_{E_i^h}X_j^h,\ldots,X_p^h)
\di\right),\\Q_2&=&-2\di\sum_{i=1}^{2n}
\sum_{l<j}T(X_1^h,\ldots,D_{E_i^h}X_l^h,\ldots,D_{E_i^h}X_j^h,\ldots,X_p^h),\\
Q_3&=& -N.N.T(X_1^h,\ldots,X_p^h)
\di+2\sum_{j=1}^pN.T(X_1^h,\ldots,D_{N}X_j^h,\ldots,X_p^h)\\
&&+D_{N}N.T(X_1^h,\ldots,X_p^h)
\di-\sum_{j=1}^pT(X_1^h,\ldots,D_{D_{N}N}X_j^h,\ldots,X_p^h)\\
&&\di-\sum_{j=1}^pT(X_1^h,\ldots,D_{N}D_{N}X_j^h,\ldots,X_p^h)
-2\sum_{l<j}T(X_1^h,\ldots,D_{N}X_l^h,\ldots,D_{N}X_j^h,\ldots,X_p^h),\\
Q_4&=& -J_0N.J_0N.T(X_1^h,\ldots,X_p^h)
\di+2\sum_{j=1}^pJ_0N.T(X_1^h,\ldots,D_{J_0N}X_j^h,\ldots,X_p^h)\\
&&+D_{J_0N}J_0N.T(X_1^h,\ldots,X_p^h)
\di-\sum_{j=1}^pT(X_1^h,\ldots,D_{D_{J_0N}J_0N}X_j^h,\ldots,X_p^h)\\
&&\di-\sum_{j=1}^pT(X_1^h,\ldots,D_{J_0N}D_{J_0N}X_j^h,\ldots,X_p^h)\\
&&-2\sum_{l<j}T(X_1^h,\ldots,D_{J_0N}X_l^h,\ldots,D_{J_0N}X_j^h,\ldots,X_p^h).\end{eqnarray*}
By using $(4)-(8)$, we get
\begin{eqnarray}D_{D_{E_i^h}E_i^h}X_j^h&=&(\na_{\na_{ E_i} E_i} X_j)^h+
<(\na_{ E_i} E_i)^h,(J X_j)^h>J_0N- <(\na_{ E_i} E_i)^h,
X_j^h>N\nonumber\\&&-
[N,X_j]-X_j,\\
D_{E_i}D_{E_i}X_j&=& (\na_{ E_i}\na_{ E_i} X_j)^h+< E_i^h,(\na_{
E_i}J X_j)^h>J_0N
-< E_i^h,(\na_{ E_i} X_j)^h>N\nonumber\\
&&+E_i^h.<E_i^h,(JX_j)^h>J_0N
-E_i^h.<E_i^h,X_j^h>N\nonumber\\&&+<E_i^h,(JX_j)^h>J_0E_i^h-<E_i^h,X_j^h>E_i^h,
\\D_ND_NX_j^h&=&[N,[N,X_j^h]]+\frac2r[N,X_j^h]+(\frac1{r^2}-\frac1r)(X_j^h-<X_j^h,N>N)\nonumber
\\&&-\frac2rN.<X_j^h,N>N,\\
D_{J_0N}D_{J_0N}X_j^h&=&[J_0N,[J_0N,X_j^h]]+\frac2r[J_0N,J_0X_j^h]-\frac1{r^2}X_j^h\nonumber\\&&-
\frac2r<D_{J_0N}X_j^h,N>J_0N.\end{eqnarray}

A careful verification using $(8)-(10)$ leads to
\begin{eqnarray*} Q_1&=&\sum_{i=1}^{2n}\left(- E_i E_i.\phi(T)(
X_1,\ldots, X_p)
+2\sum_{j=1}^p E_i.\phi(T)( X_1,\ldots,\na_{ E_i} X_j,\ldots, X_p)\right.\\
&&+\na_{ E_i} E_i.\phi(T)( X_1,\ldots, X_p)
-\sum_{j=1}^p\phi(T)( X_1,\ldots,\na_{\na_{ E_i} E_i} X_j,\ldots, X_p)\\
&&\left.-\sum_{j=1}^p\phi(T)( X_1,\ldots,\na_{ E_i}\na_{ E_i}
X_j,\ldots, X_p) \di\right)\\&&+2p(n+1)\phi(T)( X_1,\ldots,
X_p)\di -2nL_NT(X_1^h,\ldots,X_p^h)\\&&\di
+2\sum_{j=1}^p\left(J_0X_j^h.T(X_1^h,\ldots,
\stackrel{j}{\overbrace{J_0N}},\ldots,X_p^h)
-X_j^h.T(X_1^h,\ldots,\stackrel{j}{\overbrace{N}},\ldots,X_p^h)\right).\end{eqnarray*}

Now, by using $(8)$, we get $$\begin{array}{l}
Q_2=\di-2\sum_{l<j}\sum_{i=1}^{2n} \phi(T)(X_1,\ldots,\na_{
E_i} X_l,\ldots,\na_{ E_i} X_j,\ldots, X_p)\\
\di+2\sum_{l<j}
T(X_1^h,\ldots,\stackrel{l}{\overbrace{N}},\ldots,D_{X_l^h}X_j^h,\ldots,X_p^h)
+2\sum_{l<j}T(X_1^h,\ldots,D_{X_j^h}X_l^h,\ldots,\stackrel{j}{\overbrace{N}},
\ldots,X_p^h)
\\\di-2\sum_{l<j}
T(X_1^h,\ldots,\stackrel{l}{\overbrace{J_0N}},\ldots,D_{J_0X_l^h}X_j^h,\ldots,X_p^h)
-2\sum_{l<j}T(X_1^h,\ldots,D_{J_0X_j^h}X_l^h,\ldots,
\stackrel{j}{\overbrace{J_0N}},\ldots,X_p^h)\\
\di+2
\sum_{l<j}<X_l^h,X_j^h>\left(T(X_1^h,\ldots,\stackrel{l}{\overbrace{N}},\ldots,
\stackrel{j}{\overbrace{N}},\ldots,X_p^h)
+T(X_1^h,\ldots,\stackrel{l}{\overbrace{J_0N}},\ldots,
\stackrel{j}{\overbrace{J_0N}},\ldots,X_p^h)\right)
\\\di
-2
\sum_{l<j}<J_0X_l^h,X_j^h>\left(T(X_1^h,\ldots,\stackrel{l}{\overbrace{J_0N}},\ldots,
\stackrel{j}{\overbrace{N}},\ldots,X_p^h)
-T(X_1^h,\ldots,\stackrel{l}{\overbrace{N}},\ldots,
\stackrel{j}{\overbrace{J_0N}},\ldots,X_p^h)\right).\end{array}$$

We deduce that
\begin{eqnarray*}
Q_1+Q_2&=&\na^*\na\phi(T)( X_1,\ldots, X_p)+2p(n+1)\phi(T)(
X_1,\ldots,
X_p)-2nL_NT(X_1^h,\ldots,X_p^h)\\
&&+2\sum_{j=1}^p\left(D_{J_0X_j^h}(i_{J_0N,j}T)(X_1^h,\ldots,
\hat{X_j^h},\ldots,X_p^h) -D_{X_j^h}(i_{N,j}T)(X_1^h,\ldots,
\hat{X_j^h},\ldots,X_p^h)\right)
\end{eqnarray*}$$\begin{array}{l}\di+2
\sum_{l<j}<X_l^h,X_j^h>\left(T(X_1^h,\ldots,\stackrel{l}{\overbrace{N}},\ldots,
\stackrel{j}{\overbrace{N}},\ldots,X_p^h)
+T(X_1^h,\ldots,\stackrel{l}{\overbrace{J_0N}},\ldots,
\stackrel{j}{\overbrace{J_0N}},\ldots,X_p^h)\right)
\\\di
+2
\sum_{l<j}\Om_0(X_l^h,X_j^h)\left(T(X_1^h,\ldots,\stackrel{l}{\overbrace{N}},\ldots,
\stackrel{j}{\overbrace{J_0N}},\ldots,X_p^h)
+T(X_1^h,\ldots,\stackrel{l}{\overbrace{J_0N}},\ldots,
\stackrel{j}{\overbrace{J_0J_0N}},\ldots,X_p^h)\right).\end{array}$$

Remark that $\De_{P^n(\comp)}\phi(T)=\na^*\na\phi(T)+R(\phi(T))$
where $R(\phi(T))$ is given by Lemma 2.2. We deduce hence
\begin{eqnarray}
&&Q_1+Q_2-\De_{P^n(\comp)}\phi(T)(X_1,\ldots,X_p)=-2nL_NT(X_1^h,\ldots,X_p^h)+4T^{J_0}
(X_1^h,\ldots,X_p^h)\nonumber\\
&&+2\overline{T}^{J^0}(X_1^h,\ldots,X_p^h)-
2\sum_{\sigma\in\T}T^\sigma(X_1^h,\ldots,X_p^h)\nonumber\\
&&+2\sum_{j=1}^p\left(D_{J_0X_j^h}(i_{J_0N,j}T)(X_1^h,\ldots,
\hat{X_j^h},\ldots,X_p^h) -D_{X_j^h}(i_{N,j}T)(X_1^h,\ldots,
\hat{X_j^h},\ldots,X_p^h)\right)\nonumber\\
&&+2\sum_{i<j}<X_i^h,X_j^h>{\mathrm{Tr}}_{i,j}T(X_1^h,\ldots,\hat
X_i^h,\ldots,\hat X_j^h,\ldots,X_p^h)\nonumber\\
&&+2\sum_{i<j}\Om_0(X_i^h,X_j^h){\mathrm{Tr}}_{i,j,J_0}T(X_1^h,\ldots,\hat
X_i^h,\ldots,\hat X_j^h,\ldots,X_p^h).\end{eqnarray}

Let us compute $Q_3$.  Now by using $(5)$ and $(11)$ and by taking
the restriction to $S^{2n+1}$, we have
\begin{eqnarray*}
2\sum_{j=1}^pN.T(X_1^h,\ldots,D_{N}X_j^h,\ldots,X_p^h)&=&\di
2\sum_{j=1}^pN.T(X_1^h,\ldots,[{N},X_j^h],\ldots,X_p^h)\\&& +
2\sum_{j=1}^pN(\frac1r)T(X_1^h,\ldots,X_j^h,\ldots,X_p^h)\\&&\di
+2\sum_{j=1}^pN.T(X_1^h,\ldots,X_j^h,\ldots,X_p^h)\\&&\di
-2\sum_{j=1}^pN(<X_j^h,N>)T(X_1^h,\ldots,\stackrel{j}{\overbrace{N}},\ldots,X_p^h)\\\di
&=&2\sum_{j=1}^pN.T(X_1^h,\ldots,[{N},X_j^h],\ldots,X_p^h)\\&&
-2pT(X_1^h,\ldots,X_p^h)+2pN.T(X_1^h,\ldots,X_p^h)\\&&\di
-2\sum_{j=1}^pN(<X_j^h,N>)T(X_1^h,\ldots,\stackrel{j}{\overbrace{N}},\ldots,X_p^h).\\\di
\sum_{j=1}^pT(X_1^h,\ldots,D_{N}D_{N}X_j^h,\ldots,X^h_p)&=&2\sum_{j=1}^p
T(X_1^h,\ldots,[{N},X_j^h],\ldots,X_p^h)\\&&\di
\sum_{j=1}^pT(X_1^h,\ldots,[N,[N,X_j^h],\ldots,X_p^h)\\&&
-2\sum_{j=1}^pN(<X_j^h,N>)T(X_1^h,\ldots,\stackrel{j}{\overbrace{N}},\ldots,X_p^h).
\end{eqnarray*}
$$\begin{array}{l}\di
\sum_{i<j}T(X_1^h,\ldots,D_{N}X_i^h,\ldots,D_{N}X_j^h,\ldots,X_p^h)=\\\di
\sum_{i<j}T(X_1^h,\ldots,[{N},X_i^h],\ldots,[{N},X_j^h],\ldots,X_p^h)
+\frac{p(p-1)}2T(X_1^h,\ldots,X_p^h)\\\di
+\sum_{i<j}T(X_1^h,\ldots,X_i^h,\ldots,[{N},X_j^h],\ldots,X_p^h)\\\di+
\sum_{i<j}T(X_1^h,\ldots,[{N},X_i^h],\ldots,X_j^h,\ldots,X_p^h)=\\
\di\sum_{i<j}T(X_1^h,\ldots,[{N},X_i^h],\ldots,[{N},X_j^h],\ldots,X_p^h)
+\frac{p(p-1)}2T(X_1^h,\ldots,X_p^h)\\\di
+(p-1)\sum_{j=1}^pT(X_1^h,\ldots,[{N},X_j^h],\ldots,X_p^h).\end{array}$$

So we get, in restriction to $S^{2n+1}$,
\begin{equation}
Q_3=\di -L_N\circ
L_NT(X_1^h,\ldots,X_p^h)+2pL_NT(X_1^h,\ldots,X_p^h)-p(1+p)T(X_1^h,\ldots,X_p^h).
\end{equation}

Let us compute $Q_4$. By using $(7)$ et $(12)$, we get in
restriction to $S^{2n+1}$,
\begin{eqnarray*}Q_4&=&-J_0N.J_0N.T(X_1^h,\ldots,X_p^h)
+2\sum_{j=1}^pJ_0N.T(X_1^h,\ldots,[J_0N,X_j^h],\ldots,X_p^h)\\
&&+2\sum_{j=1}^pJ_0N.T(X_1^h,\ldots,J_0X_j^h,\ldots,X_p^h)\\
&&-N.T(X_1^h,\ldots,X_p^h)+\sum_{j=1}^pT(X_1^h,\ldots,[N,X_j^h],\ldots,X_p^h)+
pT(X_1^h,\ldots,X_p^h)\\
&&-\sum_{j=1}^pT(X_1^h,\ldots,[J_0N,[J_0N,X_j^h]],\ldots,X_p^h)
-2\sum_{j=1}^pT(X_1^h,\ldots,[J_0N,J_0X_j^h],\ldots,X_p^h)\\&&+pT(X_1^h,\ldots,X_p^h)
-2\sum_{i<j}T(X_1^h,\ldots,[J_0N,X_i^h],\ldots,[J_0N,X_j^h],\ldots,X_p^h)\\
&&-2\sum_{i<j}T(X_1^h,\ldots,[J_0N,X_i^h],\ldots,J_0X_j^h,\ldots,X_p^h)\\
&&-2\sum_{i<j}T(X_1^h,\ldots,J_0X_i^h,\ldots,[J_0N,X_j^h],\ldots,X_p^h)\\
&&-2\sum_{i<j}T(X_1^h,\ldots,J_0X_i^h,\ldots,J_0X_j^h,\ldots,X_p^h).\end{eqnarray*}
Hence \begin{eqnarray}Q_4 &=&-L_{J_0N}\circ
L_{J_0N}T(X_1^h,\ldots,X_p^h)-L_{N}T(X_1^h,\ldots,X_p^h)
+2pT(X_1^h,\ldots,X_p^h)\nonumber\\&&+2\sum_{j=1}^pL_{J_0N}T(X_1^h,\ldots,J_0X_j^h,\ldots,X_p^h)
-2T^{J_0}(X_1^h,\ldots,X_p^h).
\end{eqnarray}

Note that
\begin{eqnarray*}
\phi(\De_{\comp^{n+1}}T)(X_1,\ldots,X_p)-\De_{P^n(\comp)}\phi(T)(X_1,\ldots,X_p)&=&
Q_1+Q_2+Q_3+Q_4\\&&-\De_{P^n(\comp)}\phi(T)(X_1,\ldots,X_p)\end{eqnarray*}and
we get the desired formula by using $(13)-(15)$, by noting that
$L_{J_0N_0}T=0$ and by remarking that the following formulas holds
in restriction to $S^{2n+1}$
\begin{eqnarray*}
\sum_{j=1}^pD_{J_0X_j^h}(i_{J_0N,j}T)(
X_1^h,\ldots,\hat{X^h}_j,\ldots,X_p^h)&=&
\sum_{j=1}^pD_{J_0X_j^h}(i_{J_0\overrightarrow{r},j}T)(
X_1^h,\ldots,\hat{X}_j^h,\ldots,X_p^h),\\
\sum_{j=1}^pD_{X_j^h}(i_{N,j}T)(
X_1^h,\ldots,\hat{X}_j^h,\ldots,X_p^h)&=&
\sum_{j=1}^pD_{X_j^h}(i_{\overrightarrow{r},j}T)(
X_1^h,\ldots,\hat{X}_j^h,\ldots,X_p^h),\end{eqnarray*}
$$L_NT=L_{\overrightarrow{r}}T\qquad\mbox{and}\quad
L_N\circ L_NT=-L_{\overrightarrow{r}}T+L_{\overrightarrow{r}}\circ
L_{\overrightarrow{r}}T.$$ $\Box$

\section{Spectra and symmetric eigentensors of the Lichnerowicz Laplacian on
$P^n(\comp)$}

In this section, we formulate  Theorem 2.1 in the context of
symmetric covariant tensor fields (Theorem 3.1), we adapt to our
context the results obtained in [11] in the context of
differential forms and we introduce an algebraic lemma (Lemma
3.3). Hereafter, we deduce from this lemma and  Theorem 3.1 the
mains results of this paper, namely, Theorems 3.2 and 3.3.
Finally, we tabulate the eigenvalues and the eigenspaces of
$\De_{P^n(\comp)}$ acting on  differential 1-forms and on
symmetric covariant 2-tensor fields (Tables II-VIII).\bigskip

The following result is an immediate consequence of Theorem 2.1,
Remark 2.1 and the definitions of $W$, $\overline{W}$, $\de^*_h$
and $\overline{\de^*_h}$.

\begin{th} Let $T$ be a symmetric $p$-tensor field on $\comp^{n+1}$ such that
$L_ZT=0$. Then
\begin{eqnarray*}
\phi(\De_{\comp^{n+1}}T)-\De_{P^n(\comp)}\phi(T)&=&\phi\left(2p(1-p)T+2(p-n)
L_{\overrightarrow{r}}T-L_{\overrightarrow{r}}\circ
L_{\overrightarrow{r}}T\right.\\
&&\left.+4T^{J_0}-4\overline{\de^*_h}i_{\overline{W}}T-4{\de^*_h}i_{W}T+2<\;,\;>\odot
{\mathrm{Tr}}T\right).\end{eqnarray*}
\end{th}

The results on harmonic homogeneous forms on $\comp^{n+1}$,
obtained by Ikeda and Taniguchi in [11], can be adapted easily to
get similar results on harmonic homogeneous symmetric covariant
tensor fields.

Let $\s P_{k,l}^{p,q}$ be the set of symmetric tensor field $T$ on
$\comp^{n+1}$ of the form
$$T=\sum_{\stackrel{0\leq i_1<\ldots,<i_p\leq n}{0\leq j_1<\ldots,<j_q\leq
n}}T_{i_1,\ldots,i_p,j_1,\ldots,j_q}dz_{i_1}\odot\ldots\odot
dz_{i_p}\odot d\bar z_{j_1}\odot\ldots\odot d\bar z_{j_q},$$where
$T_{i_1,\ldots,i_p,j_1,\ldots,j_q}$ are polynomials of degree $k$
with respect $z_0,\ldots,z_n$ and of degree $l$ with respect $\bar
z_0,\ldots,\bar z_n$. Put
$$\s H_{k,l}^{p,q}=\s P_{k,l}^{p,q}\cap \ker\de\cap
\ker\De_{\comp^{n+1}}\quad\mbox{and}\quad T_{k,l}^{p,q}=\s
H_{k,l}^{p,q}\cap \ker{\mathrm{Tr}}.$$ In the same way as [11,
Lemma 6.4], we have
\begin{equation}\s P_{k,l}^{p,q}=\s
H_{k,l}^{p,q}\oplus\left(W^*\odot \s
P_{k,l-1}^{p-1,q}+\overline{W}^*\odot \s P_{k-1,l}^{p,q-1} +r^2\s
P_{k-1,l-1}^{p,q}\right).\end{equation}

Furthermore, in the same way as [11, Corollary 7.11], we get the
following Lemma.
\begin{Le}
$\di\phi:\bigoplus_{\stackrel{0\leq
m\leq\min(p,q)}{k+p=l+q}}<\;,\;>^m\odot
T_{k,l}^{p-m,q-m}\too\s^{p,q}(P^n(\comp),\comp)$ is injective and
its image is dense.\end{Le}

The following lemma can be obtained easily by a direct
computation.
\begin{Le} For $T\in \s P_{k,l}^{p,q}$, we have
\begin{enumerate}\item $i_W\de^*_hT-\de^*_hi_WT=(k-p)T$;
\item
$i_{\overline{W}}\overline{\de^*_h}T-\overline{\de^*_h}i_{\overline{W}}T=(l-q)T;$
\item $i_{\overline{W}}\de^*_hT-\de^*_hi_{\overline{W}}T=0$;\item
$i_W\overline{\de^*_h}T-\overline{\de^*_h}i_WT=0.$\end{enumerate}\end{Le}

Note that  the operators $i_W,i_{\overline{W}},\de^*_h$ and
$\overline{\de^*_h}$ preserve the spaces $T_{k,l}^{p,q}$, namely,
\begin{equation}\begin{array}{ccc}i_W:T_{k,l}^{p,q}\too
T_{k+1,l}^{p-1,q}&,&i_{\overline{W}}:T_{k,l}^{p,q}\too
T_{k,l+1}^{p,q-1},\\
\de^*_h:T_{k,l}^{p,q}\too
T_{k-1,l}^{p+1,q}&,&\overline{\de^*_h}:T_{k,l}^{p,q}\too
T_{k,l-1}^{p,q+1}.\end{array}\end{equation}

The task is now to decompose $T_{k,l}^{p,q}$ as a direct sum of
spaces whose the images by $\phi$ are eigenspaces of
$\De_{P^n(\comp)}$ and get, according to Lemma 3.1 and $(2)$, all
the eigenvalues. This decomposition is based on the following
algebraic lemma.

\begin{Le} Let $V$ be a finite dimensional vectorial space,
$\phi$ and $\psi$ are two endomorphisms of $V$ and
$(A_k^p)_{k,p\in\nat\cup\{-1\}}$ is a family of vectorial
subspaces of $V$ such that:
\begin{enumerate}\item for any $p,k\in\nat$,
$A^p_{-1}=A_k^{-1}=\{0\}$; \item for any $p,k\in\nat$,
$\phi(A_k^p)\subset A_{k-1}^{p+1}$ and $\psi(A_k^p)\subset
A_{k+1}^{p-1}$;\item for any $p,k\in\nat$ and for any $a\in
A_k^p$,
$$\phi\circ\psi(a)-\psi\circ\phi(a)=(p-k)a.$$\end{enumerate}
Then:

$(i)$ for any $k<p$, $\psi:A_k^p\too A_{k+1}^{p-1}$ is injective;

$(ii)$ for any $k\leq p$, we have

$$A_k^p=(A_k^p\cap \ker\phi)\oplus \psi(A_{k-1}^{p+1})\quad
\mbox{and}\quad
A_k^p=\bigoplus_{l=0}^k\psi^l(A_{k-l}^{p+l}\cap\ker\phi).$$
\end{Le}

{\bf Proof.} Note that one can deduce easily, by induction, that
for any $l\in\nat$ and for any $a\in A_k^p$
\begin{eqnarray}
\phi^l\circ\psi(a)-\psi\circ\phi^l(a)&=&l(p-k+l-1)\phi^{l-1}(a),\\
\psi^l\circ\phi(a)-\phi\circ\psi^l(a)&=&l(k-p+l-1)\psi^{l-1}(a).\end{eqnarray}

$(i)$ Let $a\in A_k^p$ such that $\psi(a)=0$. From $(18)$ and
since $p-k>0$, for any $l\geq0$, if $\phi^l(a)=0$ then
$\phi^{l-1}(a)=0$. Now, since $\phi^l(a)\in A^{p+l}_{k-l}$ and
since $A^{p+l}_{-1}=0$, we have that for any $l\geq k+1$
$\phi^l(a)=0$ which implies, by induction, that $a=0$ and hence
$\psi:A_k^p\too A_{k+1}^{p-1}$ is injective.

$(ii)$ Suppose that $k\leq p$. We define $P_k^p:A_k^p\too A_k^p$
as follows
$$\left\{\begin{array}{lll}
P_k^p(a)=&\di\sum_{s=0}^k\al_s\psi^s\circ\phi^s(a)&\\
\al_0=1&\mbox{and}\;\al_s-(s+1)(k-p-s-2)\al_{s+1}=0&\;\mbox{for}\;
1\leq s\leq k-1.\end{array}\right.$$ $P_k^p$ satisfies
$$P_k^p\circ P_k^p=P_k^p,\quad \ker P_k^p=\psi(A_{k-1}^{p+1})\quad\mbox{and}\quad
ImP_k^p=A_k^p\cap \ker\phi.$$ Indeed, let $a\in A_{k-1}^{p+1}$. We
have
\begin{eqnarray*}
P_k^p(\psi(a))&=&\sum_{s=0}^k\al_s\psi^s\circ\phi^s(\psi(a))\\
&\stackrel{(18)}=&\sum_{s=0}^k\al_s\psi^{s+1}\circ\phi^s(a)+
\sum_{s=0}^ks(p-k+s+1)\al_s\psi^s\circ\phi^{s-1}(a)\\
&\stackrel{\phi^k(a)=0}=&\sum_{s=0}^{k-1}\al_s\psi^{s+1}\circ\phi^s(a)+
\sum_{s=1}^ks(p-k+s+1)\al_s\psi^s\circ\phi^{s-1}(a)\\
&=&\sum_{s=0}^{k-1}(\al_s+(s+1)(p-k+s+2)\al_{s+1})\psi^{s+1}\circ\phi^s(a)\\
&=&0.\end{eqnarray*}

Conversely, since
$P_k^p(a)=a+\sum_{s=1}^k\al_s\psi^s\circ\phi^s(a)$, we deduce that
$P_k^p(a)=0$ implies that $a\in\psi(A_{k-1}^{p+1})$, so we have
shown that $\ker P_k^p=\psi(A_{k-1}^{p+1})$. The relation
$P_k^p\circ P_k^p=P_k^p$ is a consequence of the definition of
$P_k^p$ and $P_k^p\circ\psi=0.$

Note that $\phi(a)=0$ implies that $P_k^p(a)=a$ and hence
$A_k^p\cap\ker\phi\subset ImP_k^p$. Conversely, let $a\in A_k^p$,
we have
\begin{eqnarray*}
\phi\circ P_k^p(a)&=&\sum_{s=0}^k\al_s\phi\circ\psi^s\circ\phi^s(a)\\
&\stackrel{(19)}=&\sum_{s=0}^k\al_s\psi^s\circ\phi^{s+1}(a)-
\sum_{s=0}^k\al_ss(k-p-s-1)\psi^{s-1}\circ\phi^s(a)\\
&\stackrel{\phi^{k+1}(a)=0}=&
\sum_{s=0}^{k-1}\al_s\psi^s\circ\phi^{s+1}(a)-
\sum_{s=1}^k\al_ss(k-p-s-1)\psi^{s-1}\circ\phi^s(a)\\
&=&\sum_{s=0}^{k-1}(\al_s-(s+1)(k-p-s-2)\al_{s+1})\psi^s\circ\phi^{s+1}(a)\\
&=&0.
\end{eqnarray*}
We conclude that $P_k^p$ is a projector, $\ker
P_k^p=\psi(A_{k-1}^{p+1})$ and $A_k^p\cap\ker\phi= ImP_k^p$ and we
deduce immediately that $A_k^p=\psi(A_{k-1}^{p+1})\oplus
A_k^p\cap\ker\phi$. The same decomposition holds for
$A_{k-1}^{p+1}$, and since $\psi:A_{k-1}^{p+1}\too A_{k}^{p}$ is
injective, we get
$$A_k^p=\psi\circ\psi(A_{k-2}^{p+2})\oplus\psi(A_{k-1}^{p+1}\cap\ker\phi)\oplus
A_k^p\cap\ker\phi.$$We proceed by induction and we get the desired
decomposition.

$\Box$\bigskip

Let us apply this lemma to the operators $(i_W,\de^*_h)$ and
$(i_{\overline{W}},\overline{\de^*_h})$ acting on   the spaces
$T_{k,l}^{p,q}$.

Indeed, from Lemma 3.2 and  $(17)$ we deduce that, for  $q,l$
fixed, the spaces $(T_{k,l}^{p,q})_{k,p}$ and the operators
$(i_W,\de^*_h)$ satisfy the hypothesis  of Lemma 3.3. Thus we have
$$T_{k,l}^{p,q}=\left\{\begin{array}{ccc}
\di\bigoplus_{r=0}^k(i_{W})^r\left(T_{k-r,l}^{p+r,q}\cap
\ker\de^*_h\right)&\mbox{if}& k\leq p,\\
\di\bigoplus_{r=0}^p(\de^*_h)^r\left(T_{k+r,l}^{p-r,q}\cap \ker
i_W\right)&\mbox{if}& k\geq p.\end{array}\right.$$

On other hand, Lemma 3.2 and $(3)$ imply that the operators
$\overline{\de^*_h}$ and $i_{\overline{W}}$ commute with $\de_h^*$
and $i_W$. Hence, by Lemma 3.2 and $(17)$, for  $p,q,r$ fixed,
$((i_{W})^r\left(T_{k-r,l}^{p+r,q}\cap
\ker\de^*_h\right),\overline{\de^*_h},i_{\overline{W}})_{q,l}$ and
$((\de^*_h)^r\left(T_{k+r,l}^{p-r,q}\cap \ker
i_W\right),\overline{\de^*_h},i_{\overline{W}})_{q,l}$ satisfy
also the hypothesis of Lemma 3.3. Thus we get the following direct
sum decompositions of $T_{k,l}^{p,q}$.
\begin{equation}T_{k,l}^{p,q}=\left\{\begin{array}{ccc}
\di\bigoplus_{\stackrel{r=0,\ldots,k}{s=0,\ldots,l}} (i_{\overline
W})^s\circ
(i_W)^r\left(T_{k-r,l-s}^{p+r,q+s}\cap\ker\de^*_h\cap\ker\overline{\de_h^*}\right)&
\mbox{if}&k\leq p,l\leq q,\\
\di\bigoplus_{\stackrel{r=0,\ldots,k}{s=0,\ldots,q}}
(\overline{\de^*_h})^s\circ
(i_W)^r\left(T_{k-r,l+s}^{p+r,q-s}\cap\ker\de^*_h\cap\ker
i_{\overline{W}}\right)& \mbox{if}&k\leq p,q\leq l,\\
\di\bigoplus_{\stackrel{r=0,\ldots,p}{s=0,\ldots,l}}(i_{\overline
W})^s\circ (\de^*_h)^r\left(T_{k+r,l-s}^{p-r,q+s}\cap \ker
i_W\cap\ker\overline{\de_h^*}\right)&\mbox{if}& p\leq k,l\leq q,\\
\di\bigoplus_{\stackrel{r=0,\ldots,p}{s=0,\ldots,q}}(\overline{\de^*_h})^s\circ
(\de^*_h)^r\left(T_{k+r,l+s}^{p-r,q-s}\cap \ker i_W\cap\ker
i_{\overline{W}}\right)&\mbox{if}& p\leq k,q\leq l.
\end{array}\right.\end{equation}

Note that one can deduce easily, by applying twice the first
relation in  Lemma 3.3 $(ii)$, that if $k\leq p$ and $l\leq q$
then
\begin{equation}\dim\left(T_{k,l}^{p,q}\cap\ker\de^*_h\cap\ker\overline{\de_h^*}\right)=
\dim T_{k,l}^{p,q}+\dim T_{k-1,l-1}^{p+1,q+1} -\dim
T_{k-1,l}^{p+1,q}-\dim T_{k,l-1}^{p,q+1}.\end{equation}

Note also that since the operators $\de^*_h$ and $i_W$ and the
operators $\overline{\de_h^*}$ and $i_{\overline{W}}$ play
symmetric roles, we have
\begin{eqnarray}
\dim(T_{k,l}^{p,q}\cap\ker\de^*_h\cap\ker i_{\overline{W}})&=&
\dim(T_{k,q}^{p,l}\cap\ker\de^*_h\cap\ker\overline{\de_h^*}),\quad k\leq p,q\leq l\nonumber\\
\dim(T_{k,l}^{p,q}\cap \ker i_W\cap\ker\overline{\de_h^*})&=&
\dim(T_{p,l}^{k,q}\cap\ker\de^*_h\cap\ker\overline{\de_h^*}),\quad p\leq k,l\leq q\nonumber \\
\dim(T_{k,l}^{p,q}\cap \ker i_W\cap\ker i_{\overline{W}})&=&
\dim(T_{p,q}^{k,l}\cap\ker\de^*_h\cap\ker\overline{\de_h^*}),\quad
p\leq k,q\leq l\nonumber.\\\;\end{eqnarray}

 The next step is to show that
the image by $\phi$ of the spaces composing the direct sum
decompositions $(20)$ are actually eigenspaces of
$\De_{P^n(\comp)}$. For simplicity, we put
\begin{eqnarray*}
E_{k,l,m}^{p,q}(W^r,\overline{W}^s)&=&<\;,\;>^m\odot\left((i_{\overline
W})^s\circ
(i_W)^r\left(T_{k-r,l-s}^{p+r,q+s}\cap\ker\de^*_h\cap\ker\overline{\de_h^*}\right)\right),\\
E_{k,l,m}^{p,q}(W^r,(\overline{\de^*_h})^s)&=&<\;,\;>^m\odot\left((\overline{\de^*_h})^s\circ
(i_W)^r\left(T_{k-r,l+s}^{p+r,q-s}\cap\ker\de^*_h\cap\ker
i_{\overline{W}}\right)\right),\\
E_{k,l,m}^{p,q}((\de^*_h)^r,\overline{W}^s)&=&<\;,\;>^m\odot\left((i_{\overline
W})^s\circ (\de^*_h)^r\left(T_{k+r,l-s}^{p-r,q+s}\cap \ker
i_W\cap\ker\overline{\de_h^*}\right)\right),\\
E_{k,l,m}^{p,q}((\de^*_h)^r,(\overline{\de^*_h})^s)&=&<\;,\;>^m\odot\left((\overline{\de^*_h})^s\circ
(\de^*_h)^r\left(T_{k+r,l+s}^{p-r,q-s}\cap \ker i_W\cap\ker
i_{\overline{W}}\right)\right).\end{eqnarray*}

\begin{Le} Let $k+p=q+l$ and $T\in
{E_{k,l,m}^{p,q}(W^r,\overline{W}^s)}\cup
{E_{k,l,m}^{p,q}(W^r,(\overline{\de^*_h})^s)}\cup
{E_{k,l,m}^{p,q}((\de^*_h)^r,\overline{W}^s)}\cup
{E_{k,l,m}^{p,q}((\de^*_h)^r,(\overline{\de^*_h})^s)}$. Then
\begin{eqnarray*}\De_{P^n(\comp)}\phi(T)&=&4\left((p+k)(n-q+k)+p(p-1)+q(q-1)+r(r+1)+s(s+1)\right.\\
&&\left.+r|p-k|+s|q-l|+\frac12\left(|p-k|+|q-l|+p-k+q-l\right)\right)\phi(T).\\
\end{eqnarray*}\end{Le}

{\bf Proof.} Remark that, since $\phi(<\;,\;>\odot
T)=g\odot\phi(T)$ and by $(2)$, it suffices to show the lemma for
$T\in (i_{\overline W})^s\circ
(i_W)^r\left(T_{k-r,l-s}^{p+r,q+s}\cap\ker\de^*_h\cap\ker\overline{\de_h^*}\right)
\cup(\overline{\de^*_h})^s\circ
(i_W)^r\left(T_{k-r,l+s}^{p+r,q-s}\cap\ker\de^*_h\cap\ker
i_{\overline{W}}\right)\cup (i_{\overline W})^s\circ
(\de^*_h)^r\left(T_{k+r,l-s}^{p-r,q+s}\cap \ker
i_W\cap\ker\overline{\de_h^*}\right)\cup
(\overline{\de^*_h})^s\circ
(\de^*_h)^r\left(T_{k+r,l+s}^{p-r,q-s}\cap \ker i_W\cap\ker
i_{\overline{W}}\right)$.

Now, since $\De_{\comp^{n+1}}T=0$, we will deduce the lemma  by
computing  the right side of the formula composing Theorem 3.1.

 First note that  ${\mathrm{Tr}}T=0$,
$L_{\overrightarrow{r}}T=(p+q+k+l)T=2(p+k)T$ and
$T^{J_0}=\frac12((p+q)-(p-q)^2)T$. Thus
\begin{eqnarray*}2(p+q)(p+q-1)T+2(n-p-q)
L_{\overrightarrow{r}}T+L_{\overrightarrow{r}}\circ
L_{\overrightarrow{r}}T-4T^{J_0}=\\
(4(p+k)(n-q+k)+4(p(p-1)+q(q-1)))T.\end{eqnarray*}

Now let us compute  $\de^*_hi_WT$ and
$\overline{\de^*_h}i_{\overline{W}}T$. The computation is based on
$(18)$ applied to the operators $(\de^*_h,i_W)$ and
$(\overline{\de_h^*},i_{\overline{W}})$.

Let $T=i_{\overline{W}^s}\circ i_{W^r}(T')$ with $T'\in
T_{k-r,l-s}^{p+r,q+s}\cap\ker\de^*_h\cap\ker\overline{\de_h^*}$.
We have
\begin{eqnarray*} -\de_h^*\circ i_{W}\circ i_{\overline{W}^s}\circ
i_{W^r}(T')&=&-\de_h^*\circ i_{W^{r+1}}\circ i_{\overline{W}^s}(T')\\
&\stackrel{(18)}=&(r+1)(k-p-r)T.
\\-\overline{\de_h^*}\circ i_{\overline{W}}\circ i_{\overline{W}^s}\circ
i_{W^r}(T')&=&-\overline{\de_h^*}\circ i_{\overline{W}^{s+1}}\circ
i_{W^r}(T')\\
&\stackrel{(18)}=&(s+1)(l-q-s)T.
\end{eqnarray*}

In a similar fashion, we get:\begin{enumerate}

\item for $T=(\overline{\de^*_h})^s\circ i_{W^r}(T')$ with $T'\in
T_{k-r,l+s}^{p+r,q-s}\cap\ker\de^*_h\cap\ker i_{\overline{W}}$,
\begin{eqnarray*} -\de_h^*\circ i_{W}\circ(\overline{\de^*_h})^s\circ
i_{W^r}(T')
&=&(r+1)(k-p-r)T\\
-\overline{\de_h^*}\circ
i_{\overline{W}}\circ(\overline{\de^*_h})^s\circ
i_{W^r}(T')&=&s(q-l-s-1)T;
\end{eqnarray*}

\item for $T=i_{\overline{W}^s}\circ (\de^*_h)^r(T')$ with $T'\in
T_{k+r,l-s}^{p-r,q+s}\cap \ker i_W\cap\ker\overline{\de_h^*}$,
\begin{eqnarray*} -\de_h^*\circ i_{W}\circ i_{\overline{W}^s}\circ
(\de^*_h)^r(T')
&=&r(p-k-r-1)T.\\
-\overline{\de_h^*}\circ i_{\overline{W}}\circ
i_{\overline{W}^s}\circ (\de^*_h)^r(T')
&=&(s+1)(l-q-s)T;\end{eqnarray*}

\item for $T=(\overline{\de^*_h})^s\circ (\de^*_h)^r(T')$ with
$T'\in T_{k+r,l+s}^{p-r,q-s}\cap \ker i_W\cap\ker
i_{\overline{W}}$,
\begin{eqnarray*} -\de_h^*\circ i_{W}\circ (\overline{\de^*_h})^s\circ
(\de^*_h)^r(T')&=&r(p-k-r-1)T.\\
-\overline{\de_h^*}\circ i_{\overline{W}}\circ
(\overline{\de^*_h})^s\circ
(\de^*_h)^r(T')&=&s(q-l-s-1)T.\end{eqnarray*}
\end{enumerate} The lemma follows by gathering together  all the results above.$\Box$

\bigskip

It is natural now to compute the multiplicities of the eigenvalues
obtained in Lemma 3.4. So, by Lemma 3.1, we need to compute the
dimension of the spaces ${E_{k,l,m}^{p,q}(W^r,\overline{W}^s)}$,
${E_{k,l,m}^{p,q}(W^r,(\overline{\de^*_h})^s)}$,
${E_{k,l,m}^{p,q}((\de^*_h)^r,\overline{W}^s)}$ and
${E_{k,l,m}^{p,q}((\de^*_h)^r,(\overline{\de^*_h})^s)}$. Note
first that, according to  Lemma 3.3 $(i)$, we have
\begin{eqnarray*}\dim {E_{k,l,m}^{p,q}(W^r,\overline{W}^s)}&=&
T_{k-r,l-s}^{p+r,q+s}\cap\ker\de^*_h\cap\ker\overline{\de_h^*},\quad
k\leq p,l\leq q,\\
\dim {E_{k,l,m}^{p,q}(W^r,(\overline{\de^*_h})^s)}&=&
T_{k-r,l+s}^{p+r,q-s}\cap\ker\de^*_h\cap\ker
i_{\overline{W}},\quad k\leq
p,q\leq l,\\
\dim {E_{k,l,m}^{p,q}((\de^*_h)^r,\overline{W}^s)}&=&
T_{k+r,l-s}^{p-r,q+s}\cap\ker i_W\cap\ker \overline{\de_h^*},\quad
p\leq
k,l\leq q,\\
\dim {E_{k,l,m}^{p,q}((\de^*_h)^r,(\overline{\de^*_h})^s)}&=&
T_{k+r,l+s}^{p-r,q-s}\cap\ker i_W\cap\ker i_{\overline{W}},\quad
p\leq
k,q\leq l.\\
\end{eqnarray*}

From $(22)$, to get the multiplicities it suffices to compute the
dimension of
$T_{k,l}^{p,q}\cap\ker\de^*_h\cap\ker\overline{\de_h^*}$ for any
$k\leq p$ and $l\leq q$. According to $(21)$, this will be done if
one compute the dimension of $T_{k,l}^{p,q}$ for any $p,q,k,l$.

Since
$$\s H_{k,l}^{p,q}=T_{k,l}^{p,q}\oplus <\;,\;>\odot \s
H_{k,l}^{p-1,q-1},$$we have
\begin{equation}\dim T_{k,l}^{p,q}=\dim\s H_{k,l}^{p,q}-\dim\s
H_{k,l}^{p-1,q-1}.\end{equation}

To conclude, we need the following lemma.

\begin{Le} We have
\begin{eqnarray*}
\dim\s H_{k,l}^{p,q}&=&\dim\s P_{k,l}^{p,q}+\dim\s
P_{k-1,l-1}^{p-1,q-1}+\dim \s P_{k-1,l-2}^{p-1,q}+\dim\s
P_{k-2,l-1}^{p,q-1}\\
&&-\left(\dim\s P_{k,l-1}^{p-1,q}+\dim\s P_{k-1,l}^{p,q-1}+\dim \s
P_{k-1,l-1}^{p,q}+\dim\s
P_{k-2,l-2}^{p-1,q-1}\right).\end{eqnarray*}
\end{Le}

{\bf Proof.} The  relation is a consequence of $(16)$ and the
following equalities
$$\begin{array}{l} (W^*\odot \s
P_{k,l-1}^{p-1,q})\cap(\overline{W}^*\odot \s P_{k-1,l}^{p,q-1})=W
^*\odot\overline{W}^*\odot \s
P_{k-1,l-1}^{p-1,q-1}.\\
\left(W^*\odot \s P_{k,l-1}^{p-1,q}+\overline{W}^*\odot \s
P_{k-1,l}^{p,q-1}\right)\cap r^2\s P_{k-1,l-1}^{p,q}= r^2\left(
W^*\odot \s P_{k-1,l-2}^{p-1,q}+\overline{W}^*\odot \s
P_{k-2,l-1}^{p,q-1}\right).
\end{array}$$$\Box$

\bigskip

Thus, from Lemma 3.5, $(23)$ and $(21)$, and after many
simplifications, we get, for any $k\leq p$ and for any $l\leq q$,
\begin{eqnarray}
\dim\left(
T_{k,l}^{p,q}\cap\ker\de^*_h\cap\ker\overline{\de_h^*}\right)&=&\dim\s
P_{k,l}^{p,q}+\dim\s P_{k,l-1}^{p-2,q-1}+\dim\s
P_{k-1,l}^{p-1,q-2}\nonumber\\&&+\dim\s P_{k-2,l-2}^{p-2,q-2}
+\dim\s P_{k-1,l-1}^{p+1,q+1}+\dim \s
P_{k-2,l-3}^{p,q+1}\nonumber\\&&+\dim\s P_{k-3,l-2}^{p+1,q}+\dim\s
P_{k-3,l-3}^{p-1,q-1}+\dim\s P_{k-2,l}^{p+1,q-1}\nonumber\\&&
+\dim\s P_{k-3,l-1}^{p,q-2}+\dim\s P_{k,l-2}^{p-1,q+1}+\dim \s
P_{k-1,l-3}^{p-2,q}\nonumber\\&& -\dim\s P_{k,l}^{p-1,q-1}-\dim\s
P_{k-1,l-1}^{p-2,q-2} -\dim \s
P_{k-2,l-2}^{p+1,q+1}\nonumber\\&&-\dim\s P_{k-3,l-3}^{p,q}
-\dim\s P_{k-1,l}^{p+1,q}-\dim\s
P_{k-3,l-1}^{p+1,q-1}\nonumber\\
&&-\dim\s P_{k-2,l}^{p,q-2}-\dim\s P_{k-3,l-2}^{p-1,q-2} -\dim\s
P_{k,l-1}^{p,q+1}\nonumber\\&&-\dim \s P_{k-1,l-3}^{p-1,q+1}
-\dim\s P_{k,l-2}^{p-2,q}-\dim\s
P_{k-2,l-3}^{p-2,q-1}\nonumber.\end{eqnarray}

Even if this formula involves a great deal of terms, it is
surprising that after a computation using computing software  and
the formula
$$\dim_{\comp} \s P_{k,l}^{p,q}=\left(\begin{array}{c}n+k
\\k
\end{array}\right) \left(\begin{array}{c} n+l\\l
\end{array}\right) \left(\begin{array}{c} n+p\\p
\end{array}\right) \left(\begin{array}{c} n+q\\q
\end{array}\right),$$where
$$\left(\begin{array}{c}a
\\b
\end{array}\right)=\frac{a!}{b!(a-b)!},$$ we get a simple expression of
$\dim\left(
T_{k,l}^{p,q}\cap\ker\de^*_h\cap\ker\overline{\de_h^*}\right)$.
 Let us tabulate the results.\bigskip

\begin{tabular}{|l|l|l|}
\hline &&\\
Conditions on &Space&Complex dimension\\
$p,q,k,l$&&\\ \hline&&\\  $1\leq k\leq p,2\leq l\leq
q$&$T_{k,l}^{p,q}\cap\ker\de^*_h\cap\ker\overline{\de_h^*}$&
{\footnotesize$
\frac{(n+p-2)!(n+q-2)!(n+k-3)!(n+l-3)!n^3(n-1)^2(n-2)}{(n!)^4(p+1)!(q+1)!k!l!}\times$}\\
or&&{\tiny$(p-k+1)(q-l+1)(n+k+l-2)(n+q+k-1)(n+p+l-1)\times$}\\
$2\leq k\leq p,1\leq l\leq
q$&&{\tiny$(n+p+q)$}\\\hline&&\\
$1\leq  p,1\leq
q$&$T_{1,1}^{p,q}\cap\ker\de^*_h\cap\ker\overline{\de_h^*}$&
{\footnotesize$
\frac{(n+p-2)!(n+q-2)!n^2(n-2)pq(n+q)(n+p)(n+p+q)}{(n!)^2(p+1)!(q+1)!}$}\\
&&\\ \hline&&\\
 $1\leq p,1\leq l\leq
q$&$T_{0,l}^{p,q}\cap\ker\de^*_h\cap\ker\overline{\de_h^*}$&
{\footnotesize$
\frac{(n+p-2)!(n+q-1)!(n+l-2)!n^2(n-1)(q-l+1)(n+p+l-1)(n+p+q)}{(n!)^3p!(q+1)!l!}$}\\
&&\\\hline&&\\
 $1\leq k\leq p,1\leq
q$&$T_{k,0}^{p,q}\cap\ker\de^*_h\cap\ker\overline{\de_h^*}$&
{\footnotesize$
\frac{(n+p-1)!(n+q-2)!(n+k-2)!n^2(n-1)(p-k+1)(n+q+k-1)(n+p+q)}{(n!)^3q!(p+1)!k!}$}\\
&&\\
\hline&&\\
 $1\leq p,1\leq
q$&$T_{0,0}^{p,q}\cap\ker\de^*_h\cap\ker\overline{\de_h^*}$& {$
\frac{(n+p-1)!(n+q-1)!n(n+p+q)}{(n!)^2p!q!}$}\\
&&\\\hline&&\\  $0\leq l\leq
q$&$T_{0,l}^{0,q}\cap\ker\de^*_h\cap\ker\overline{\de_h^*}$& {$
\frac{(n+q)!(n+l-1)!n(q-l+1)}{2(n!)^2(q+1)!l!}$}\\&&\\ \hline&&\\
$0\leq k\leq
p$&$T_{k,0}^{p,0}\cap\ker\de^*_h\cap\ker\overline{\de_h^*}$& {$
\frac{(n+p)!(n+k-1)!n(p-k+1)}{2(n!)^2(p+1)!k!}$}\\&&\\ \hline
\end{tabular}
\begin{center} {\footnotesize Table I}.\end{center}\bigskip

We are now able to  give the spectra and the eigenspaces with
multiplicities of $\De_{P^n(\comp)}$ acting on
$\s^{p,q}(P^n(\comp),\comp)$. Note that the spectra of
$\De_{P^n(\comp)}$ acting on $\s^{p,q}(P^n(\comp),\comp)$ is the
same as the spectra of $\De_{P^n(\comp)}$ acting on
$\s^{q,p}(P^n(\comp),\comp)$ and the eigenspaces are conjugated.
So, we restrict ourself to the case $p\leq q$.

Fix $p,l\in\nat$ and consider the space
$\s^{p,p+l}(P^n(\comp),\comp)$. We have, from Lemma 3.1,
$$\di\phi:\bigoplus_{\stackrel{0\leq m\leq
p}{k\in\nat}}<\;,\;>^m\odot
T_{k+l,k}^{p-m,p+l-m}\too\s^{p,p+l}(P^n(\comp),\comp)$$ is
injective and its image is dense. To obtain the eigenvalues and
eigenspaces of $\De_{P^n(\comp)}$ acting on
$\s^{p,p+l}(P^n(\comp),\comp)$, we split  any
$T_{k+l,k}^{p-m,p+l-m}$ above according to $(20)$ and we apply
Lemma 3.4. Note also that, according to Table I, the dimension of
some eigenspaces vanishes when $n=1$ or $n=2$ and so, one must
distingue three cases $n\geq3$, $n=2$ or $n=1$. To state the
results, we introduce some notations.

We put
\begin{eqnarray*}S_0&=&\left\{(m,k,r,s)\in\nat^4/0\leq m\leq
p,0\leq k<p-m-l,0\leq r\leq k+l,0\leq s\leq
k\right\},\\
S_1&=&\left\{(m,k,r,s)\in\nat^4/0\leq m\leq p,\max(0,p-m-l)\leq
k<p-m+l,0\leq r\leq p-m,\right.\\&&\left.0\leq s\leq k\right\},\\
S_2&=&\left\{(m,k,r,s)\in\nat^4/0\leq m\leq p,k\geq p-m+l,0\leq
r\leq p-m,0\leq s\leq p-m+l\right\},\\
V_{r,s,0}^{p,l,m,k}&=&{E_{k+l,k,m}^{p-m,p+l-m}}(W^r,\overline{W}^s)\quad
\mbox{if}\quad (m,k,r,s)\in S_0,\\
V_{r,s,1}^{p,l,m,k}&=&{E_{k+l,k,m}^{p-m,p+l-m}}((\de_h^*)^r,\overline{W}^s)
\quad \mbox{if}\quad (m,k,r,s)\in S_1,\\
V_{r,s,2}^{p,l,m,k}&=&{E_{k+l,k,m}^{p-m,p+l-m}}((\de_h^*)^r,(\overline{\de^*_h})^s)\quad
\mbox{if}\quad (m,k,r,s)\in S_2.\\\end{eqnarray*} The eigenvalue
obtained in Lemma 3.4 becomes
$$\begin{array}{l} \la_{r,s,n}^{p,l,m,k}=
4\left[(p-m+k+l)(n-p+m+k)+(p-m)(p-m-1)\right.\\+(p-m+l)(p-m+l-1)+r(r+1)+s(s+1)
+r|p-m-k-l|+s|p-m+l-k|\\\left.+\frac12\left(|p-m-k-l|+|p-m+l-k|+2(p-m-k)\right)\right].
\end{array}$$

Finally, the following notations are needed to treat the case
$n=2$.
\begin{eqnarray*}S_0^0&=&\left\{(m,k,r)\in\nat^3/0\leq m\leq
p,0\leq k<p-m-l,0\leq r\leq
k\right\},\\
S_0^1&=&\left\{(m,k,r)\in\nat^3/0\leq m\leq p,0\leq
k<p-m-l,0\leq r\leq k+l\right\},\\
S_1^0&=&\left\{(m,k,r)\in\nat^3/0\leq m\leq p,\max(0,p-m-l)\leq
k<p-m+l,0\leq r\leq
k\right\},\\S_1^1&=&\left\{(m,k,r)\in\nat^3/0\leq m\leq
p,\max(0,p-m-l)\leq
k<p-m+l,0\leq r\leq p-m\right\},\\
S_2^0&=&\left\{(m,k,r)\in\nat^3/0\leq m\leq p,k\geq p-m+l,0\leq r\leq p-m+l\right\},\\
S_2^1&=&\left\{(m,k,r)\in\nat^3/0\leq m\leq p,k\geq p-m+l,0\leq
r\leq p-m\right\},\end{eqnarray*}\begin{eqnarray*}
W_{r,0,0}^{p,l,m,k}&=& V_{k+l,r,0}^{p,l,m,k}\quad\quad
\mbox{if}\quad (m,k,r)\in S_0^0,\quad W_{r,0,1}^{p,l,m,k}=
V_{r,k,0}^{p,l,m,k}\quad\quad
\mbox{if}\quad (m,k,r)\in S_0^1,\\
W_{r,1,0}^{p,l,m,k}&=&V_{p-m,r,1}^{p,l,m,k}\quad\quad
\mbox{if}\quad (m,k,r)\in S_1^0,\quad
W_{r,1,1}^{p,l,m,k}=V_{r,k,1}^{p,l,m,k}\quad\quad \mbox{if}\quad (m,k,r)\in S_1^1,\\
W_{r,2,0}^{p,l,m,k}&=&V_{p-m,r,2}^{p,l,m,k}\quad\quad
\mbox{if}\quad (m,k,r)\in S_2^0,\quad
W_{r,2,1}^{p,l,m,k}=V_{r,p-m+l,2}^{p,l,m,k}\quad\quad
\mbox{if}\quad (m,k,r)\in S_2^1.\\\end{eqnarray*}

The following theorem is an immediate consequence of Lemma 3.1,
(20) and Lemma 3.4.
\begin{th} Let $n,p,l\in\nat$ such that $n\geq3$. Then:
\begin{enumerate}\item $\di\phi:\bigoplus_{i=0,\ldots,2}\left(
\bigoplus_{(m,k,r,s)\in
S_i}V_{r,s,i}^{p,l,m,k}\right)\too\s^{p,p+l}(P^n(\comp),\comp)$ is
injective and its image is dense, \item for any $i=0,\ldots,2$ and
for any $(m,k,r,s)\in S_i$, $\phi(V_{r,s,i}^{p,l,m,k})$ is an
eigenspace of $\De_{P^n(\comp)}$ associated to the eigenvalue
$\la_{r,s,n}^{p,l,m,k},$ \item for any $i=0,\ldots,2$ and for any
$(m,k,r,s)\in S_i$, the dimension of $\phi(V_{r,s,i}^{p,l,m,k})$
is given by Table I and $(22)$ since
\begin{eqnarray*}
\dim\left(\phi(V_{r,s,0}^{p,l,m,k})\right)&=&
\dim\left(T_{k+l-r,k-s}^{p-m+r,p-m+l+s}\cap\ker\de^*_h\cap\ker\overline{\de_h^*}\right),\\
\dim\left(\phi(V_{r,s,1}^{p,l,m,k})\right)&=&\dim\left(T_{k+l+r,k-s}^{p-m-r,p-m+l+s}
\cap\ker\overline{\de^*_h}\cap\ker
i_{{W}}\right),\\
\dim\left(\phi(V_{r,s,2}^{p,l,m,k})\right)&=&\dim\left(T_{k+l+r,k+s}^{p-m-r,p-m+l-s}\cap\ker
i_W\cap\ker i_{\overline{W}}\right).\end{eqnarray*}
\end{enumerate}\end{th}

By deleting in Theorem 4.2 the spaces $V_{r,s,i}^{p,l,m,k}$ whose
dimension vanishes in the case $n=2$ or $n=1$, we get the
following theorem.

\begin{th} Let $p,l\in\nat$. Then:
\begin{enumerate}\item $\di\phi:\bigoplus_{i,j=0,\ldots,2}\left(
\bigoplus_{(m,k,r)\in
S_i^j}W_{r,i,j}^{p,l,m,k}\right)\too\s^{p,p+l}(P^2(\comp),\comp)$
is injective and its image is dense; \item the spaces
$W_{r,0,0}^{p,l,m,k}$, $W_{r,0,1}^{p,l,m,k}$,
$W_{r,1,0}^{p,l,m,k}$, $W_{r,1,1}^{p,l,m,k}$,
$W_{r,2,0}^{p,l,m,k}$ and $W_{r,2,1}^{p,l,m,k}$ are eigenspaces of
$\De_{P^2(\comp)}$ associated, respectively, to the eigenvalues
$\la_{k+l,r,2}^{p,l,m,k}$, $\la_{r,k,2}^{p,l,m,k}$,
$\la_{p-m,r,2}^{p,l,m,k}$, $\la_{r,k,2}^{p,l,m,k}$,
$\la_{p-m,r,2}^{p,l,m,k}$, $\la_{r,p-m+l,2}^{p,l,m,k}$; \item  the
dimension of $\phi(W_{r,i,j}^{p,l,m,k})$ is given by Table I and
$(22)$ since
\begin{eqnarray*}
\dim\left(\phi(W_{r,0,0}^{p,l,m,k})\right)&=&
\dim\left(T_{0,k-r}^{p-m+k+l,p-m+l+r}\cap\ker\de^*_h\cap\ker\overline{\de_h^*}\right),
\\
\dim\left(\phi(W_{r,0,1}^{p,l,m,k})\right)&=&
\dim\left(T_{k+l-r,0}^{p-m+r,p-m+l+k}\cap\ker\de^*_h\cap\ker\overline{\de_h^*}\right),\\
\dim\left(\phi(W_{r,1,0}^{p,l,m,k})\right)&=&\dim\left(T_{k+l+p-m,k-r}^{0,p-m+l+r}
\cap\ker\overline{\de^*_h}\cap\ker
i_{{W}}\right),\\
\dim\left(\phi(W_{r,1,1}^{p,l,m,k})\right)&=&\dim\left(T_{k+l+r,0}^{p-m-r,p-m+l+k}
\cap\ker\overline{\de^*_h}\cap\ker
i_{{W}}\right),\\
\dim\left(\phi(W_{r,2,1}^{p,l,m,k})\right)&=&\dim\left(T_{k+l+p-m,k+r}^{0,p-m+l-r}\cap\ker
i_W\cap\ker i_{\overline{W}}\right),\\
\dim\left(\phi(W_{r,2,2}^{p,l,m,k})\right)&=&\dim\left(T_{k+l+r,k+p-m+l}^{p-m-r,0}\cap\ker
i_W\cap\ker i_{\overline{W}}\right);\end{eqnarray*}

\item for $P^1(\comp)$, we have
\begin{eqnarray*}\phi:\bigoplus_{\stackrel{0\leq m\leq p}{0\leq
k<p-m-l}}\left(V_{k+l-1,k-1,0}^{p,l,m,k}\oplus
V_{k+l,k,0}^{p,l,m,k}\right)\oplus \bigoplus_{\stackrel{0\leq
m\leq p}{\max(0,p-m-l)\leq
k<p-m+l}}\left(V_{p-m-1,k-1,1}^{p,l,m,k}\oplus
V_{p-m,k,1}^{p,l,m,k}\right)\\\oplus \bigoplus_{\stackrel{0\leq
m\leq p}{k\geq p-m+l}}\left(V_{p-m-1,p-m+l-1,2}^{p,l,m,k}\oplus
V_{p-m,p-m+l,2}^{p,l,m,k}\right)\too\s^{p,p+l}(P^1(\comp),\comp)\end{eqnarray*}
is injective and its image is dense. Moreover, the image by $\phi$
of all the spaces $V_{r,s,i}^{p,l,m,k}$ composing the above direct
sum decomposition are eigenspaces of $\De_{P^1(\comp)}$ associated
to the eigenvalue $\la_{r,s,1}^{p,l,m,k}$, and their dimensions
can be deduced from 3. Theorem 4.2.\end{enumerate}\end{th}
\begin{rem}In [11], Ikeda
and Taniguchi computed the eigenvalues of $\De_{P^n(\comp)}$
acting on $\Om(P^n(\comp),\comp)$ and determined the spaces of
eigenforms as representation spaces, but they did not give the
multiplicities. The formula obtained in Theorem 2.1 can be used in
the case of differential forms to show that the images by $\phi$
of the spaces composing  the direct sum decomposition (7.2) in
[11] are eigenspaces of $\De_{P^n(\comp)}$. The dimensions of
these spaces can be computed in a similar way as in [12].
Unfortunately, the formulas obtained are much more complicated
than the case of the spheres. However, in [11, Theorem 7.13],
Ikeda and Taniguchi showed that these spaces  are irreducible
$SU(n+1)$-modules and they computed their highest weights. Hence,
may be, one can use the Weyl dimension formula to compute the
dimensions of these spaces and to get the multiplicities of the
eigenvalues.
\end{rem}

Finally, we apply Theorems 3.2 and 3.3 for the low values of $p$
and $l$ and we tabulate the results.\bigskip

{\tiny
\begin{tabular}{|c|c|c|c|}
\hline &&&\\
Spaces&Eigenvalues&Eigenspaces&Complex dimension\\
$n\geq1$&$k\in\nat$&&\\
\hline
&&&\\
$C^\infty(P^n(\comp))$&$4k(n+k)$&$\phi\left(T_{k,k}^{0,0}\right)$&$\frac{n(n+2k)((n+k-1)!)^2}{(n!)^2(k!)^2}$\\
\hline &$4(n+1)$&$T_{1,0}^{0,1}$&$n(n+2)$\\
$\s^{0,1}(P^n(\comp),\comp)$&$4(k+1)(n+k+2)$&$\phi\circ\overline{\de^*_h}
\left(T_{k+2,k+2}^{0,0}\right)$
&$\frac{n(n+2k+4)((n+k+1)!)^2}{(n!)^2((k+2)!)^2}$\\
&$4(k+2)(n+k+1)$&$\phi\left(T_{k+2,k+1}^{0,1}\cap\ker
i_{\overline{W}}\right)$&$\frac{(k+1)n(n-1)(n+k+2)(n+2k+3)((n+k)!)^2}{(n!)^2((k+2)!)^2}$\\
\hline
&$4(n+1)$&$\phi\left(T_{0,1}^{1,0}\right)$&$n(n+2)$\\
$\s^{1,0}(P^n(\comp),\comp)$&$4(k+2)(n+k+2)$&$\phi\circ{\de^*_h}\left(T_{k+2,k+2}^{0,0}\right)$&
$\frac{n(n+2k+4)((n+k+1)!)^2}{(n!)^2((k+2)!)^2}$\\
&$4(k+2)(n+k+1)$&$\phi\left(T_{k+1,k+2}^{1,0}\cap\ker
i_{{W}}\right)$&$\frac{(k+1)n(n-1)(n+k+2)(n+2k+3)((n+k)!)^2}{(n!)^2((k+2)!)^2}$\\
\hline
\end{tabular}}

\begin{center} {\tiny Table II}.\end{center}\bigskip

{\tiny
\begin{tabular}{|c|c|c|c|}
\hline &&&\\
Spaces&Eigenvalues&Eigenspaces&Complex dimension\\
$n\geq1$&$k\in\nat$&&\\
\hline
&&&\\
&$8(n+2)$&$\phi\left(T_{2,0}^{0,2}\cap\ker\overline{\de^*_h}\right)$&$\frac{n(n+4)(n+1)^2}4$\\
&$12(n+3)$&$\phi\circ
i_{\overline{W}}\left(T_{3,0}^{0,3}\cap\ker\overline{\de^*_h}\right)$&$
\frac{n(n+1)^2(n+2)^2(n+6)}{36}$\\
$\s^{0,2}(P^n(\comp),\comp)$&$4(k+4)(n+k+4)$&$\phi\circ(\overline{\de_h^*})^2\left(T^{0,0}_{k+4,k+4}\right)$&
$\frac{((n+k+3)!)^2n(n+2k+8)}{(n!)^2((k+4)!)^2}$\\
&$12(n+2)$&$\phi\left(T_{3,1}^{0,2}\cap\ker\overline{\de^*_h}\right)$&
$\frac{n(n+1)^2(n-1)(n+2)(n+5)}9$\\
&$4(k+4)(n+k+3)$&$\phi\circ\overline{\de_h^*}\left(T^{0,1}_{k+4,k+3}\cap\ker
i_{\overline{W}}\right)$&
$\frac{((n+k+2)!)^2n(n-1)(k+3)(n+k+4)(n+2k+7)}{(n!)^2((k+4)!)^2}$\\
&{\tiny$4(k^2+(n+6)k+4n+10)$}&$\phi\left(T^{0,2}_{k+4,k+2}\cap\ker
i_{\overline{W}}\right)$&{\tiny
$\frac{(n+k+2)!(n+k+1)!n^2(n-1)(k+1)(n+k+5)(n+2k+6)}{2(n!)^2(k+4)!(k+3)!}$}\\
\hline\end{tabular}}

\begin{center} {\tiny Table III}.\end{center}\bigskip

{\tiny
\begin{tabular}{|c|c|c|c|}
\hline &&&\\
Spaces&Eigenvalues&Eigenspaces&Complex dimension\\
$n\geq1$&$k\in\nat$&&\\
\hline
&&&\\
&$8(n+2)$&$\phi\left(T_{0,2}^{2,0}\cap\ker{\de^*_h}\right)$&$\frac{n(n+4)(n+1)^2}4$\\
&$12(n+3)$&$\phi\circ
i_{{W}}\left(T_{0,3}^{3,0}\cap\ker{\de^*_h}\right)$&$
\frac{n(n+1)^2(n+2)^2(n+6)}{36}$\\
$\s^{2,0}(P^n(\comp),\comp)$&$4(k+4)(n+k+4)$&$\phi\circ({\de_h^*})^2
\left(T^{0,0}_{k+4,k+4}\right)$&
$\frac{((n+k+3)!)^2n(n+2k+8)}{(n!)^2((k+4)!)^2}$\\
&$12(n+2)$&$\phi\left(T_{1,3}^{2,0}\cap\ker{\de^*_h}\right)$&
$\frac{n(n+1)^2(n-1)(n+2)(n+5)}9$\\
&$4(k+4)(n+k+3)$&$\phi\circ{\de_h^*}\left(T^{1,0}_{k+3,k+4}\cap\ker
i_{{W}}\right)$&
$\frac{((n+k+2)!)^2n(n-1)(k+3)(n+k+4)(n+2k+7)}{(n!)^2((k+4)!)^2}$\\
&{\tiny$4(k^2+(n+6)k+4n+10)$}&$\phi\left(T^{2,0}_{k+2,k+4}\cap\ker
i_{{W}}\right)$&{\tiny
$\frac{(n+k+2)!(n+k+1)!n^2(n-1)(k+1)(n+k+5)(n+2k+6)}{2(n!)^2(k+4)!(k+3)!}$}\\
\hline\end{tabular}}

\begin{center} {\tiny Table IV}.\end{center}\bigskip

{\tiny
\begin{tabular}{|c|c|c|c|}
\hline &&&\\
Spaces&Eigenvalues&Eigenspaces&Complex dimension\\
$n\geq1$&$k\in\nat$&&\\
\hline
&&&\\
&$4(n+1)$&$\phi\left(T_{0,0}^{1,1}\right)$&$n(n+2)$\\
&$4(k+2)(n+k+2)$&
$\phi\circ\de_h^*\circ\overline{\de_h^*}\left(T_{k+2,k+2}^{0,0}\right)$&
$\frac{((n+k+1)!)^2n(n+2k+4)}{(n!)^2((k+2)!)^2}$\\
&$4k(n+k)$&$\phi\left(<\;,\;>\odot T_{k,k}^{0,0}\right)$&
$\frac{((n+k-1)!)^2n(n+2k)}{(n!)^2(k!)^2}$\\
$\s^{1,1}(P^n(\comp),\comp)$&$4(k+2)(n+k+1)$&
$\phi\circ\overline{\de_h^*}\left(T^{1,0}_{k+1,k+2}\cap\ker
i_W\right)$&$\frac{2((n+k)!)^2n(n-1)(k+1)(n+k+1)(n+2k+3)}{(n!)^2((k+2)!)^2}$\\
&&$\oplus\phi\circ{\de_h^*}\left(T^{0,1}_{k+2,k+1}\cap\ker
i_{\overline{W}}\right)$&\\
&$4(k+2)(n+k)$&$\phi\left(T_{k+1,k+1}^{1,1}\cap\ker i_W\cap\ker
i_{\overline{W}}\right)$&
$\frac{((n+k-1)!)^2n^2(n-2)(k+1)^2(n+k+1)^2(n+2k+2)}{(n!)^2((k+2)!)^2}$\\
\hline\end{tabular}}

\begin{center} {\tiny Table V}.\end{center}\bigskip

By setting $n=2$ in Tables III-V, we get  the eigenvalues and the
eigenspaces of $\De_{P^2(\comp)}$ acting on
$\s^2(P^2(\comp),\comp)$. These  results complete the results
obtained in [23] since we give explicitly the eigenspaces. Note
that there is a misprint in [23, Table 1 pp. 227]. The degeneracy
of $\frac23\Lambda(m+1)(m+3)$ is, actually, $2(m+2)^3$ (this is
the value obtained by Warner  in [23,(6.5)]).\bigskip

{\tiny
\begin{tabular}{|c|c|c|c|}
\hline &&&\\
Spaces&Eigenvalues&Eigenspaces&Complex dimension\\
&$m\in\nat$&&\\
\hline
&&&\\
&$32$&$\phi\left(T_{2,0}^{0,2}\cap\ker\overline{\de^*_h}\right)$&$27$\\
&$60$&$\phi\circ
i_{\overline{W}}\left(T_{3,0}^{0,3}\cap\ker\overline{\de^*_h}\right)$&$
64$\\
$\s^{0,2}(P^2(\comp),\comp)$&$4(m+4)(m+6)$&
$\phi\circ(\overline{\de_h^*})^2\left(T^{0,0}_{m+4,m+4}\right)$&
$(m+5)^3$\\
&$48$&$\phi\left(T_{3,1}^{0,2}\cap\ker\overline{\de^*_h}\right)$&
$56$\\
&$4(m+4)(m+5)$&$\phi\circ\overline{\de_h^*}\left(T^{0,1}_{m+4,m+3}\cap\ker
i_{\overline{W}}\right)$&
$\frac{(m+3)(m+6)(2m+9)}{2}$\\
&{\tiny$4(m^2+8m+18)$}&$\phi\left(T^{0,2}_{m+4,m+2}\cap\ker
i_{\overline{W}}\right)$&{\tiny
$(m+1)(m+7)(m+4)$}\\
\hline\end{tabular}}

\begin{center}{\tiny Table VI}.\end{center}\bigskip

{\tiny
\begin{tabular}{|c|c|c|c|}
\hline &&&\\
Spaces&Eigenvalues&Eigenspaces&Complex dimension\\
&$m\in\nat$&&\\
\hline
&&&\\
&$32$&$\phi\left(T_{0,2}^{2,0}\cap\ker{\de^*_h}\right)$&$27$\\
&$60$&$\phi\circ
i_{{W}}\left(T_{0,3}^{3,0}\cap\ker{\de^*_h}\right)$&$
64$\\
$\s^{2,0}(P^2(\comp),\comp)$&$4(m+4)(m+6)$&
$\phi\circ({\de_h^*})^2\left(T^{0,0}_{m+4,m+4}\right)$&
$(m+5)^3$\\
&$48$&$\phi\left(T_{1,3}^{2,0}\cap\ker{\de^*_h}\right)$&
$56$\\
&$4(m+4)(m+5)$&$\phi\circ{\de_h^*}\left(T^{1,0}_{m+3,m+4}\cap\ker
i_{{W}}\right)$&
$\frac{(m+3)(m+6)(2m+9)}{2}$\\
&{\tiny$4(m^2+8m+18)$}&$\phi\left(T^{2,0}_{m+2,m+4}\cap\ker
i_{{W}}\right)$&{\tiny
$(m+1)(m+7)(m+4)$}\\
\hline\end{tabular}}

\begin{center} {\tiny Table VII.}\end{center}\bigskip

{\tiny
\begin{tabular}{|c|c|c|c|}
\hline &&&\\
Spaces&Eigenvalues&Eigenspaces&Complex dimension\\
&$m\in\nat$&&\\
\hline
&&&\\
&$12$&$\phi\left(T_{0,0}^{1,1}\right)$&$8$\\
$\s^{1,1}(P^2(\comp),\comp)$&$4(m+2)(m+4)$&
$\phi\circ\de_h^*\circ\overline{\de_h^*}\left(T_{m+2,m+2}^{0,0}\right)$&
$(m+3)^3$\\
&$4m(m+2)$&$\phi\left(<\;,\;>\odot T_{m,m}^{0,0}\right)$&
$(m+1)^3$\\
&$4(m+2)(m+3)$&
$\phi\circ\overline{\de_h^*}\left(T^{1,0}_{m+1,m+2}\cap\ker
i_W\right)$&$(m+1)(m+3)(2m+5)$\\
\hline\end{tabular}}

\begin{center}{\tiny Table VIII}.\end{center}

\bigskip

 {\bf References}\bigskip

[1] {\bf E. Bedford and T. Suwa,} Eigenvalues of Hopf manifolds,
American Mathemaical Society, Vol. {\bf 60} (1976), 259-264.

[2] {\bf B. L. Beers and R. S. Millman,} The spectra of the
Laplace-Beltrami operator on compact, semisimple Lie groups, Amer.
J. Math., {\bf 99} (4) (1975), 801-807.

[3] {\bf M. Berger and D. Ebin,} Some decompositions of the space
of symmetric tensors on Riemannian manifolds, J. Diff. Geom., {\bf
3} (1969), 379-392.

[4] {\bf M. Berger, P. Gauduchon and E. Mazet,} Le spectre d'une
vari\'et\'e riemannienne, Lecture Notes in Math., Vol {\bf 194},
Springer Verlag (1971).

[5] {\bf A. Besse,} Einstein manifolds, Springer-Verlag,
Berlin-Hiedelberg-New York (1987).

[6] {\bf  M. Boucetta ,} Spectre des Laplaciens de Lichnerowicz
sur les sph\`eres et les projectifs r\'eels, Publicacions
Matem\`atiques, Vol. {\bf 43} (1999), 451-483.

[7] {\bf  M. Boucetta ,} Spectre du Laplacien de Lichnerowicz sur
les projectifs complexes, C. R. Acad. Sci. Paris, t. 333, S\'erie
I, (2001), 571-576.

[8] {\bf  M. Boucetta ,} Spectra and symmetric eigentensors of the
Lichnerowicz Laplacian
 on $S^n$, arXiv:0704.1363v1 [math.DG].

[9] {\bf S. Gallot and D. Meyer,} Op\'erateur de courbure et
laplacien des formes diff\'erentielles d'une vari\'et\'e
riemannienne, J. Math. Pures Appl., {\bf 54} (1975), 259-289.

[10] {\bf G. W. Gibbons and M. J. Perry,} Quantizing gravitational
instantons, Nuclear Physics B, Vol. {\bf 146}, Issue I (1978),
90-108.

[11] {\bf A. Ikeda and Y. Taniguchi,} Spectra and eigenforms of
the Laplacian on $S^n$ and $P^n(\comp)$, Osaka J. Math., {\bf 15}
(3) (1978), 515-546.

[12] {\bf I. Iwasaki and K. Katase,}  On the spectra of Laplace
operator on $\wedge^*(S^n)$, Proc. Japan Acad., {\bf 55}, Ser. A
(1979), 141-145.

[13] {\bf E. Kaneda,} The spectra of 1-forms on simply connected
compact irreducible Riemannian symmetric spaces, J. Math. Kyoto
Univ., {\bf 23} (1983), 369-395 and {\bf 24} (1984), 141-162.

[14] {\bf A. L\'evy-Bruhl-Laperri\`ere,} Spectre de de Rham-Hodge
sur les formes de degr\'e 1 des sph\`eres de $\reel^n$ ($n\geq6$),
Bull. Sc. Math., $2^e$ s\'erie, {\bf 99} (1975), 213-240.

[15] {\bf A. L\'evy-Bruhl-Laperri\`ere,} Spectre de de Rham-Hodge
sur l'espace projectif  complexe, C. R. Acad. Sc. Paris {\bf 284}
S\'erie A (1977), 1265-1267.

[16] {\bf A. Lichnerowicz,} Propagateurs et commutateurs en
relativit\'e g\'en\'erale, Inst. Hautes Etude Sci. Publ. Math.,
{\bf 10} (1961).

[17] {\bf K. Mashimo,} Spectra of Laplacian on $G_2/SO(4)$, Bull.
Fac. Gen. Ed. Tokyo Univ. of Agr. and Tech. {\bf 26} (1989),
85-92.

[18] {\bf K. Mashimo,} On branching theorem of the pair
$(G_2,SU(3))$, Nihonkai Math. J., Vol. {\bf 8} No. 2 (1997),
101-107.

[19] {\bf K. Mashimo,} Spectra of the Laplacian on the Cayley
projective plane, Tsukuba J. Math., Vol. {\bf 21} No. 2 (1997),
367-396.

[20] {\bf R. Michel,} Probl\`eme d'analyse g\'eom\'etrique li\'es
\`a la conjecture de Blaschke, Bull. Soc. Math. France, {\bf 101}
(1973), 17-69.

[21] {\bf K. Pilch and N. Schellekens,} Formulas of the
eigenvalues of the Laplacian on tensor harmonics on symmetric
coset spaces, J. Math. Phys., {\bf 25} (12) (1984), 3455-3459.

[22] {\bf C. Tsukamoto, } The sepctra of the Laplace-Beltrami
operator on $SO(n+2)/SO(2)\times SO(n)$ and $Sp(n+1)/Sp(1)\times
Sp(n)$, Osaka J. Math. {\bf 18} (1981), 407-226.

[23] {\bf N. P. Warner,} The spectra of operators on $\comp P^n$,
Proc. R. Soc. Lond. A {\bf 383} (1982), 217-230.\bigskip

Mohamed Boucetta\\
Facult\'e des Sciences et Techniques \\
BP 549 Marrakech, Morocco.
\\
Email: {\it mboucetta2@yahoo.fr }

\end{document}